\RequirePackage{fix-cm}

\documentclass[twocolumn]{svjour3}
\smartqed
\usepackage{graphicx}
\usepackage{bm}
\usepackage{url}
\usepackage{color}
\usepackage{comment}
\usepackage{gensymb}
\usepackage[utf8]{inputenc}
\usepackage[T1]{fontenc}
\usepackage{longtable}
\usepackage{amsmath}
\usepackage{amssymb}
\usepackage{commath}
\usepackage{mathrsfs}
\usepackage{esint}
\usepackage{epsfig}
\usepackage{array}
\usepackage{booktabs}
\usepackage{cite}

\begin{document}

\title{Hydrodynamic stress maps on the surface of a flexible fin-like foil}

\author{Paule Dagenais \and Christof M. Aegerter}

\institute{Physik-Institut, University of Zurich, Winterthurerstrasse 190, 8057 Zurich, Switzerland}

\date{Received: date / Accepted: date}

\maketitle

\begin{abstract}

We determine the time dependence of pressure and shear stress distributions on the surface of a pitching and deforming hydrofoil from measurements of the three dimensional flow field. Period-averaged stress maps are obtained both in the presence and absence of steady flow around the foil. The velocity vector field is determined via volumetric three-component particle tracking velocimetry and subsequently inserted into the Navier-Stokes equation to calculate the total hydrodynamic stress tensor. In addition, we also present a careful error analysis of such measurements, showing that local evaluations of stress distributions are possible. The flapping foil used in the experiments is designed to allow comparison with a small trapezoidal fish fin, in terms of the scaling laws that govern the oscillatory flow regime. Unsteady Euler-Bernoulli beam theory is employed to derive instantaneous transversal force distributions on the deflecting hydrofoil from its deflection and thereby validate the spatial distributions of hydrodynamic stresses obtained from the fluid velocity field. The consistency of the force time-dependence is verified using a control volume analysis.

\end{abstract}

\section{Introduction}

The evaluation of hydrodynamic forces on the surface of submerged and deforming foils constitutes a key element in understanding how synthetic and animal fins interact with the surrounding fluid to achieve their appropriate functions of propulsion and maneuvering. Great effort is deployed by scientists to describe the motion of fish fins and explore the physics underlying their complex kinematics \cite{Blickhan1992, Stamhuis1995, Muller1997, Drucker1999, Lauder2000, Muller2000, Drucker2001, Muller2001, Nauen2002a, Nauen2002b, Drucker2005, Muller2006, Lauder2007, Muller2008, Flammang2011a, Flammang2011b, Mwaffo2017}. The desire to reproduce the elegant complexity of nature has inspired the fabrication of elaborate fins models \cite{Tangorra2007, Tangorra2010, Dewey2012, Esposito2012, Ren2016a, Ren2016b}. Countless studies have resorted to particle imaging or particle tracking velocimetry (PIV and PTV) to investigate the flow topology and hydrodynamic forces of bio-inspired or authentic fish fins \cite{Triantafyllou2004, Godoy-Diana2008, Bohl2009, Schnipper2009, Kim2010, Green2011, David2012, Marais2012, Cheng2013, Cheng2014, Shinde2014, David2017, Muir2017}. Euler-Bernoulli beam theory has also been employed to describe the fluid-structure interactions of fins and hydrofoils, addressing concepts such as foil compliance, damping effects, resonance frequency and efficiency optimization \cite{Ramananarivo2011, Kopman2013, Kancharala2014, Iosilevskii2016, Kancharala2016, Paraz2016}.

Due to the fundamental three-dimensional complexity of fin kinematics and wake structures, the empirical investigation of their hydrodynamics must escape the limitations of planar PIV/PTV \cite{Tytell2006, Tytell2008}. Given the three-dimensionality of the wake, it still remains a big challenge to determine the propulsive forces on the moving appendages. One technique consists in determining the momentum deficit in the wake, based on the vorticity field \cite{Milne-Thomson1973, Dickinson1996a, Dickinson1996b, Drucker1999, Drucker2001, Drucker2005, Wen2013}, where the concept of vortex added-mass should also be included in a complete depiction of wake momentum exchanges \cite{Dabiri2005, Dabiri2006a, Dabiri2006b, Peng2007}. An alternative approach to calculate the total hydrodynamic force relies on a momentum balance inside a control volume and was also employed in the context of flapping fin-like structures \cite{Noca1996, Unal1997, Anderson1998, vanOudheusden2006, vanOudheusden2007, Godoy-Diana2008, Hong2008, Bohl2009, Jardin2009, Ragni2009, Ragni2010, Ragni2011, Rival2011, Leftwich2012, Shinde2014, Tronchin2015}. A non-invasive method to determine the three-dimensional pressure field inside an unsteady flow domain consists in solving the Navier-Stokes equation, in either one of the two following forms: (1) the pressure gradient is expressed in terms of spatial and temporal derivatives of the velocity field and engaged in a direct spatial integration to compute the pressure values everywhere in the domain \cite{Jakobsen1999, Baur1999, Liu2006, Murai2007, Liu2008, vanOudheusden2008, Charonko2010, deKat2010, Panciroli2013, Dabiri2014, Joshi2014, Gemmell2015, Tronchin2015, Lucas2017, McClure2017, Mwaffo2017}, or (2) the Laplacian of the pressure is obtained by taking the divergence of the latter pressure gradient formulation, and this so-called Poisson equation is solved numerically \cite{Gresho1987, Gurka1999, Fujisawa2006, Murai2007, Fujisawa2008, Windsor2008, Lorenzoni2009, Charonko2010, deKat2010, Khodarahmi2010, Koschatzky2010, Windsor2010, Kunze2011, Suryadi2011, Violato2011, deKat2012, Ghaemi2012, Koschatzky2012, deKat2013, McClure2017, vanderKindere2019}. In both cases, prescribing appropriate boundary conditions constitutes a crucial step in the calculation. An exhaustive review of theoretical and historical aspects of pressure calculations based on the velocity field is presented in \cite{vanOudheusden2013}.

In the present work, we employ the pressure gradient integration method and demonstrate the possibility to obtain well resolved distributions of hydrodynamic stress both spatially and temporally, on the surface of a synthetic fin whose size is of the order of 1 cm$^2$. The stresses here comprise both pressure and viscous shear stresses, as both contributions can have particular significance depending on the hydrodynamic problem under investigation. For this purpose, we apply a full volumetric, three components particle tracking velocimetry technique (3D-3C PTV) for the first time in this context. In our study, special attention is paid to the time and spatial resolutions and their experimental limitations, and we present a careful analysis of the error propagation from the tracked particles positions to the final hydrodynamic stress values. This work demonstrates the feasibility of experimentally capturing the full 3D force field in the fluid environment of a flapping flexible foil with dimensions comparable to that of a small fish, and to employ a force decomposition method to project the hydrodynamic stress tensor on the relevant morphological axes of the fin. Unsteady Euler-Bernoulli beam deflection theory is used as a complementary calculation to derive the instantaneous transversal force distributions acting on the deflecting hydrofoils, and validate the PTV-based stress results. Finally, the consistency of the time-dependent total force acting on the fin, obtained from the integration of the PTV-based stress distributions, is verified by comparing the results with those obtained from a control volume analysis. The main focus of the present study is to provide a proof of principle of this hydrodynamic stress evaluation approach in the case of a small trapezoidal flexible pitching foil, and to assess the implications of the temporal and spatial resolutions on the uncertainties of the reconstructed pressure and viscous shear stress fields. Potential applications of this method include the guided design of biomimetic propulsive appendages and the study of the relationship between form and function in fish locomotion problems.

\section{Material and Methods}
\subsection{Volumetric particle tracking velocimetry}
\label{sst:PTV}

PIV and PTV techniques have become widespread in the past decades in various hydrodynamics research fields and have been extensively described in the literature \cite{Maas1993, Dracos1996, Raffel1998, Pereira2006}. We performed volumetric three-components PTV measurements using the V3V-9800 system (TSI Incorporated, 500 Cardigan Road, Shoreview, Minnesota 55126 USA), which is well characterized in \cite{Lai2008}. This method presents the advantage of capturing the full unsteady flow field in a single recording of the whole 3D domain, and thus is preferable to more arduous implementations of scanning stereo-PIV \cite{Ragni2011, Suryadi2011, Ragni2012}, where a large number of measurement planes is required to reconstruct the whole volumetric flow field, or tomographic PIV which typically restricts the measurement region to a thin layer of only a few millimeters \cite{Elsinga2006}.

\begin{figure*}
\centering
\includegraphics[width=11cm]{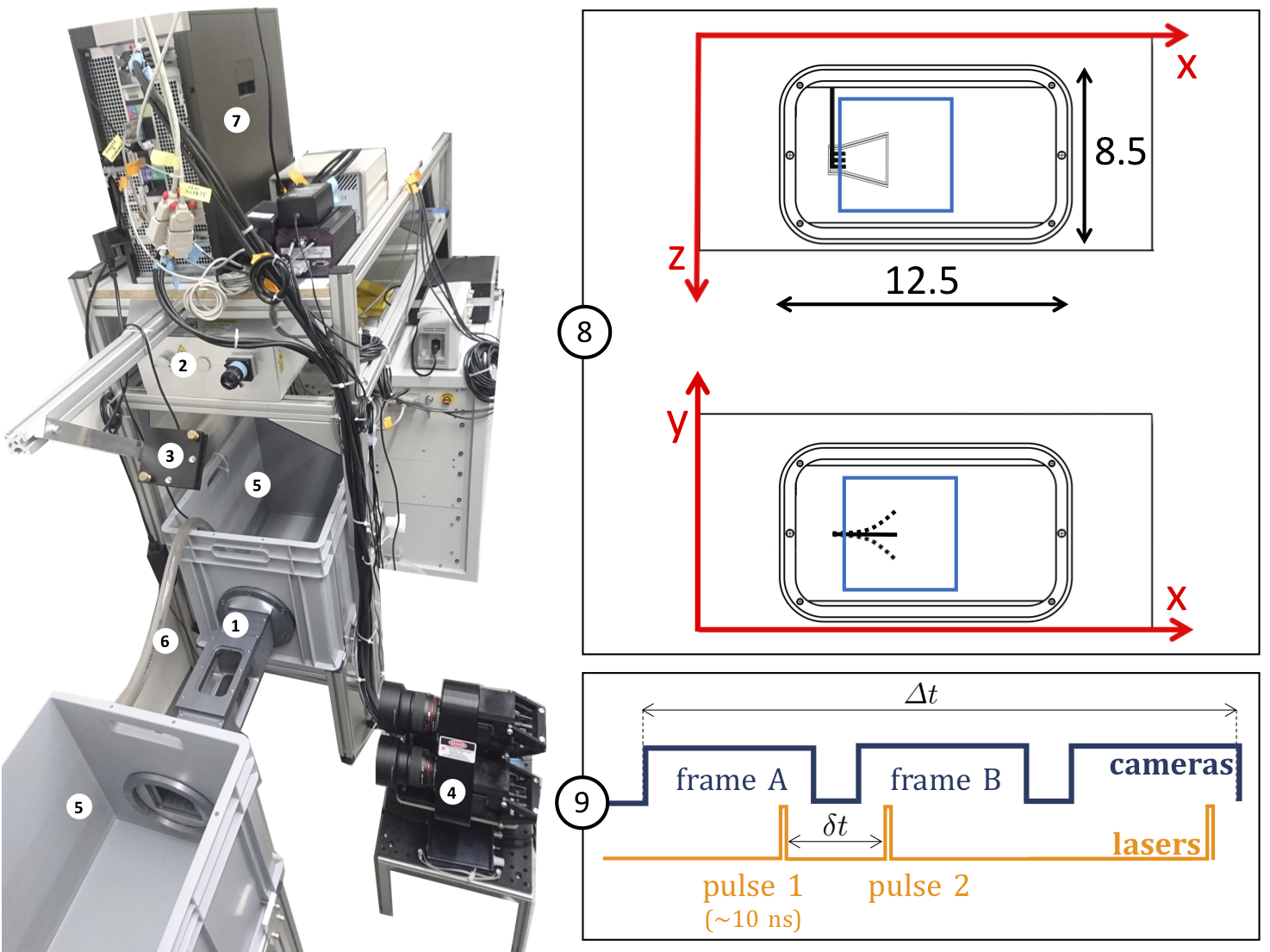}
\caption{Experimental setup for 3D-3C PTV. (1) Flow chamber with transparent windows on three sides and a fixation wall on one side for inserting the synthetic fin. (2) Dual-head pulsed Nd:YAG laser with wave-length of 532 nm and maximal energy of 120 mJ/pulse, equipped with a pair of cylindrical lenses to expand the beam. (3) Mirror to deflect the laser beam. (4) Three cameras (resolution 4 MP, 85 mm lenses, sensor size 11.3$\times$11.3 mm$^2$, magnification 0.3, maximal frequency of capture 180 Hz) mounted on a triangular plate parallel to the front wall of the flow chamber ($x$-$y$ plane). The distance between the cameras plate and the center of the water tunnel is $\sim$465 mm. (5) Water tanks for the recirculating system. (6) Pipe and pump to control the flow. (7) V3V software and synchronizer. (8) Close-up of the flow chamber with inner dimensions (in cm), PTV-interrogation volume in blue ($\sim$50$\times$50$\times$20 mm\textsuperscript{3}) and sketch of the trapezoidal fin. Frontal view ($x-y$ plane) and top view ($x-z$ plane). (9) Frame straddling method with $\delta t$ between the particle images and $\Delta t$ between the velocity fields.}
\label{f:setup}
\end{figure*}

In the present set up, illustrated in figure \ref{f:setup}, a synthetic fin is attached to a small metallic rod and inserted through the back wall of the flow chamber, with a servomotor fixed outside the water tunnel to activate the pitching program of the rotation axis. Double pulses emitted by a dual laser system illuminate a measurement volume (50$\times$50$\times$20 mm\textsuperscript{3}) inside the flow chamber. The working fluid is water and the seeding tracers are polyimide particles with a diameter of $\sim$50 $\mu$m, which were used in a precedent study of fish locomotion based on particle velocimetry \cite{Flammang2011a}. In total, $\sim$45 mL of particles are seeded in $\sim$160 L of water, which yields an approximate number of 8$\times$10\textsuperscript{4} particles inside the interrogation volume, considering that the spherical particles powder has a filling fraction of $\sim$40\%. Three cameras mounted on a triangular plate are recording triplets of images. Hence, the tracer particles advected by the flow are captured from three different perspectives at each time point. Each of the three images is analyzed based on a 2D Gaussian fit of the particle intensity distribution to identify the 2D position map. The 3D field of particle positions is reconstructed by overlapping the three viewing angles based on a triplet search algorithm. Flow velocities are determined based on individual particle displacements in the time between subsequent laser pulses ($\delta t = 2.5$ ms), using a relaxation method to achieve a probability-based matching. Appropriate synchronization is arranged between the cameras and lasers in a temporal scheme called frame straddling, as illustrated in figure \ref{f:setup}. The rate at which velocity fields are recorded (80 Hz) is half the camera acquisition frequency, yielding a time separation of $\Delta t$=0.0125 s between the velocity fields. This allows the reconstruction of the material acceleration field, necessary for the hydrodynamic stress calculation.

The image post-processing was conducted using the V3V software (version 2.0.2.7) along with an optimal set of parameters, selected to reduce the number of velocity outliers while preserving the main flow features. The maximum overlap between two particle intensity distributions was set to 65\%, a mask was applied over the foil during the particle identification step, and appropriate median filter and velocity range ($\pm$0.25 m/s) were applied to the raw velocity fields. Finally, the particle velocities were interpolated on a regular 3D grid using Gaussian interpolation, yielding a final spatial resolution of 0.75 mm in each direction. Smoothing was subsequently performed on the interpolated velocity field. The grid points inside the hydrofoil boundary, where no particle is detected owing to the mask, are not attributed a velocity value and are not involved in the hydrodynamic stress calculation (see section \ref{sst:surface_param}).

\subsection{Geometry and kinematics of the synthetic fin}

\begin{figure}
\centering
\includegraphics[width=3cm, trim={0cm 0cm 0cm 0cm}]{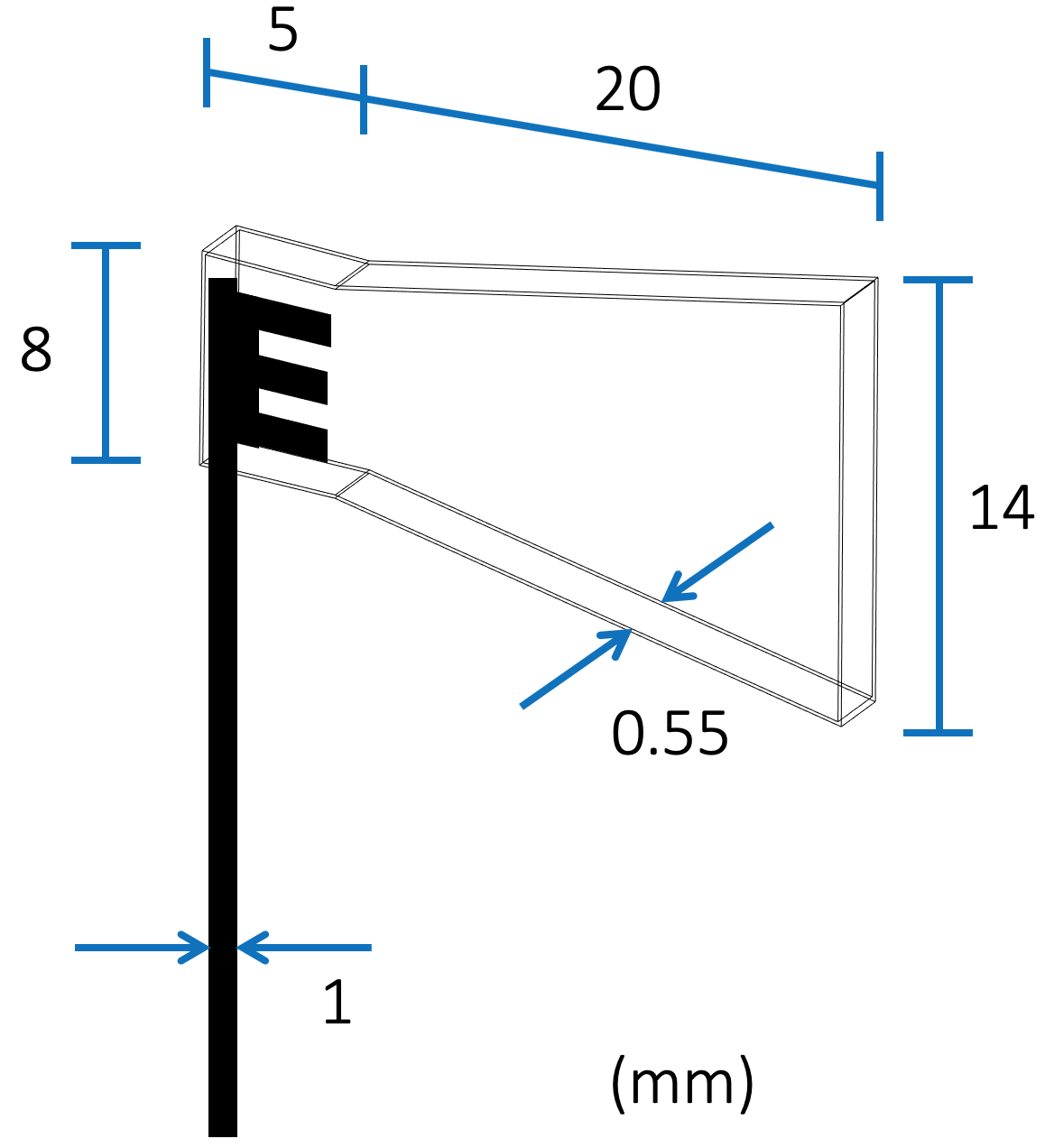}
\includegraphics[width=5cm, trim={0cm -2.5cm 2cm 2.5cm}]{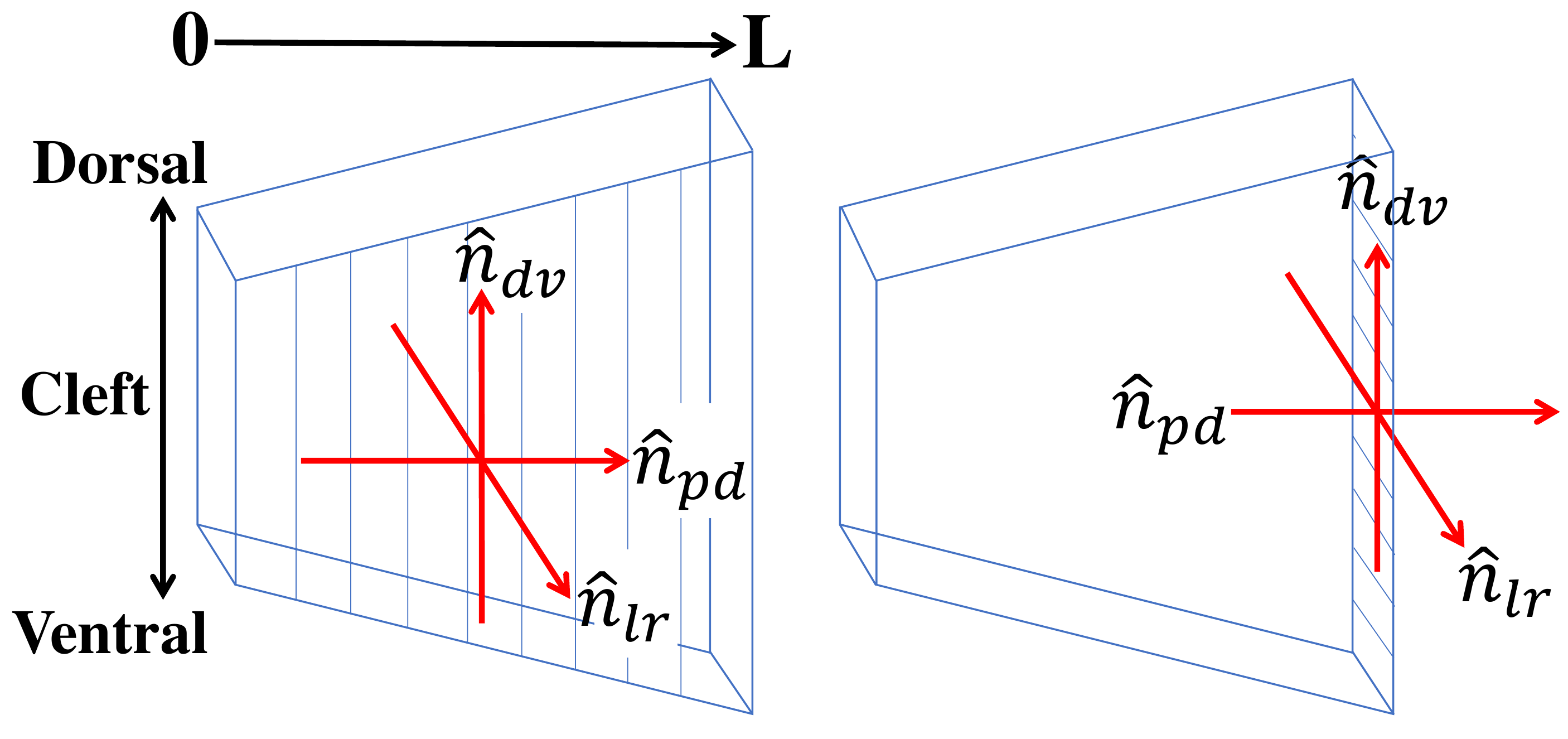}
\caption{\textit{Left}: Synthetic fin dimensions. \textit{Right}: Main axes on the trapezoidal fin (proximo-distal, dorso-ventral, left-right).}
\label{f:fin}
\end{figure}

A synthetic trapezoidal fin was designed with proportions reminiscent of the caudal fin of a small fish. The dimensions of the fin model are shown in figure \ref{f:fin} and the experimental parameters are listed in table \ref{t:experiments}. The flexible foil was produced by 3D printing a rigid cast of its negative form, then pouring liquid PDMS (polydimethylsiloxane) inside the cast, installing the fork-shaped base rod at the appropriate location, and curing the model for 36 hours at 58\degree C. This resulted in a flexible membrane of thickness 0.55 mm. The mass density fo the foil $\rho_{f}$ matches that of water (\url{http://www.mit.edu/~6.777/matprops/pdms.htm}). A study of elastic properties of real caudal fins has reported an effective Young's modulus of $\sim$8 MPa for the zebrafish, including both the rays and interray tissue contributions \cite{Puri2018a, Puri2018b}. The same cantilever deflection set up was employed by Sahil Puri (University of Zurich) to characterize the elastic properties of our PDMS foil, resulting in a value of $\sim$0.8 MPa. In order to further characterize the properties of our fin model, we performed a mechanical damping analysis, where the fin was simply held from the tip at a certain amplitude and released (in air and in water). The time evolution of the tip amplitude is described by a decaying oscillation:

\begin{equation}
y(t)=e^{-\gamma t}A cos(2 \pi \omega_0 t - \phi)
\label{eq:damping}
\end{equation}

We found values of $\gamma_{air} = $1.5 s\textsuperscript{-1} and $\gamma_{water} = $5.0 s\textsuperscript{-1} for the decay rates in air and in water, respectively, and values of $\omega_{0,air} = $9.3 rad/s and $\omega_{0,water} = $2.4 rad/s for the eigenfrequencies. These results are needed for the estimation of the damping terms in the Euler-Bernoulli equation (see section \ref{sst:euler_bernoulli}).

Two versions of the pitching fin experiment were conducted, with different pitching amplitude and upstream velocity (table \ref{t:experiments}). The dimensionless Reynolds and Strouhal numbers are employed to characterize the flow regime of the flapping foil:

\begin{equation}
Re=\frac{U \cdot L}{\nu} \hspace{1cm}  St=\frac{f \cdot a}{U}
\label{eq:ReSt}
\end{equation}

These parameters are based on the kinematic viscosity of water ($\nu$), the average fluid velocity ($U=\abs{\vec{u}}_{ave}$), the foil length ($L$=25 mm), the flapping frequency ($f$= 3Hz) and the tip-to-tip excursion amplitude ($a$, obtained from the parametrized deflection profiles - see section \ref{sst:surface_param}). Even within a single fish species, a wide range of $St$ and $Re$ can be reached by caudal fin motion, depending on the developmental stage (larval, juvenile or adult) and the swimming behavior. For example, zebrafish undergoes a developmental transition which allows it to cover a large spectrum of Reynolds and Strouhal numbers throughout its lifetime, ranging from 300 to 2000 in $Re$ and from 0.3 to 2.5  in $St$, based on fin length, swimming speed, tail beat frequency and amplitude reported by numerous studies \cite{Budick2000, Muller2004, Muller2008, Parichy2009, Palstra2010, Sfakianakis2011, Singleman2014, vanLeeuwen2015, Mwaffo2017, Voesenek2018}.

\begin{table}
\begin{center}
\begin{tabular}{c c c}
& Exp. 1 & Exp. 2  \\
\cmidrule(lr){2-2}
\cmidrule(lr){3-3}
$u_\infty$ (mm/s) & 0 & 17 \\
$\theta_0$ ($\degree$) & 10 & 25 \\
$a$ (mm) & 11 & 14 \\
$\abs{\vec{u}}_{ave}$ (mm/s) & 15.4 & 29.0 \\
$\sigma_{\abs{\vec{u}}_{ave}}$ (mm/s) & 13.0 & 15.0 \\
$Re$ & 385 & 725 \\
$St$ & 2.1 & 1.5
\end{tabular}
\caption{Parameters of the oscillating fin model for the two experimental cases}
\label{t:experiments}
\end{center}
\end{table}

\subsection{Hydrodynamic stress calculation}
\label{sst:stress_calculation}

The starting point of the hydrodynamic stress calculation is the Navier-Stokes or momentum equation \cite{Aris1990, Whitaker1968}:

\begin{equation}
\rho \frac{D\vec{u}}{Dt}=\vec{\nabla} \cdot \bm{s}+\rho \vec{g}
\label{eq:Navier-Stokes}
\end{equation}

The last term on the right corresponds to any type of body force acting on the fluid, such as gravity which can be measured independently of the flow. This can be included in a static pressure term and we will omit it in the following. The left side of the equation contains the material acceleration, with the Lagrangian derivative $D/Dt$. The first term on the right side is the divergence of the total hydrodynamic stress tensor $s_{ij}=-p\delta_{ij}+\tau_{ij}$, where $p$ is the scalar pressure field and  \bm{$\tau$} the viscous stress tensor. This viscous shear stress can be expressed as:

\begin{equation}
\tau_{ij}=\mu\bigg( \frac{\partial u_i}{\partial x_j}+\frac{\partial u_j}{\partial x_i} \bigg),
\label{eq:viscous_stress}
\end{equation}

where $\mu$ is the dynamic viscosity. Computationally, $\tau_{ij}$ can be obtained directly using centered finite differences between neighboring velocity vectors:

\begin{equation}
\begin{aligned}
\tau_{ij} \simeq {} & \mu \bigg( \frac{u_i(\vec{x}+\Delta x \hat{j})-u_i(\vec{x}-\Delta x \hat{j})}{2 \Delta x} \\
&   + \frac{u_j(\vec{x}+\Delta x \hat{i})-u_j(\vec{x}-\Delta x \hat{i})}{2 \Delta x} + O(\Delta x^2) \bigg)
\label{eq:discrete_shear_stress}
\end{aligned}
\end{equation}

In order to determine the pressure from the flow fields, we first rewrite the Navier-Stokes equation to isolate the pressure gradient on one side:

\begin{equation}
\begin{aligned}
\frac{\partial p}{\partial x_i}={} & -\rho \bigg( \frac{\partial u_i}{\partial t}+u_i \frac{\partial u_i}{\partial x_i} +u_j \frac{\partial u_i}{\partial x_j} \bigg)
+\mu \bigg( \frac{\partial^2 u_i}{\partial x_i^2} + \frac{\partial^2 u_i}{\partial x_j^2}  \bigg) \\
& + \bigg(-\rho u_k \frac{\partial u_i}{\partial x_k} +\mu \frac{\partial^2 u_i}{\partial x_k^2}\bigg)
\label{eq:pressure_gradient}
\end{aligned}
\end{equation}

The pressure field can then be reconstructed by integrating this equation, with the additional requirement of specifying the pressure values at the starting points of the integration paths $p(\vec{r}_{ref})$.
%

The pressure calculation was conducted using the \textit{queen2} algorithm \cite{Dabiri2014}, which is available at \url{http://dabirilab.com/software/}. The calculation assumes zero-pressure values on the external boundary of the domain, in the undisturbed flow, and integrates the pressure gradient along eight different paths (horizontal, vertical or diagonal) originating from the outside contour, among which the median is selected for each node in the 3D domain. The object is masked to prevent any integration path crossing its boundary. This algorithm offers the advantage of reasonable computation times even for large domains, with a total number of paths equal to 8 times the domain size ($N_x \times N_y \times N_z$) per velocity field, resulting in a more than order-of-magnitude improvement in computation time compared to other existing methods such as the multi-path integration scheme described in \cite{Liu2006}. The terms containing out-of-plane derivatives of the velocity (grouped separately in equation \ref{eq:pressure_gradient}) are not included directly in the computation. Rather, the three-dimensionality is handled through a combination of integration paths originating from all the faces of the domain, combined with a median polling. Several studies have demonstrated that a global integration approach such as a Poisson equation solver or a direct integration of the pressure gradient can bear a reasonably low level of out-of-plane motion without inducing any strong deviation from the true pressure field \cite{Charonko2010, Violato2011, deKat2012, Koschatzky2012, Lucas2017}.

The central step of the pressure calculation consists in evaluating the material acceleration, which can be done in the Eulerian or the Lagrangian frame. The Eulerian frame is stationary in the laboratory, where the time derivative of the velocity is expressed as a total derivative:

\begin{equation}
\frac{D}{Dt} = \frac{\partial}{\partial t} + \frac{\partial x_i }{\partial t} \cdot \frac{\partial}{\partial x_i}  = \frac{\partial}{\partial t} + \vec{u} \cdot \vec{\nabla}
\label{eq:material_derivative}
\end{equation}

\begin{figure}
\centering
\includegraphics[width=6cm]{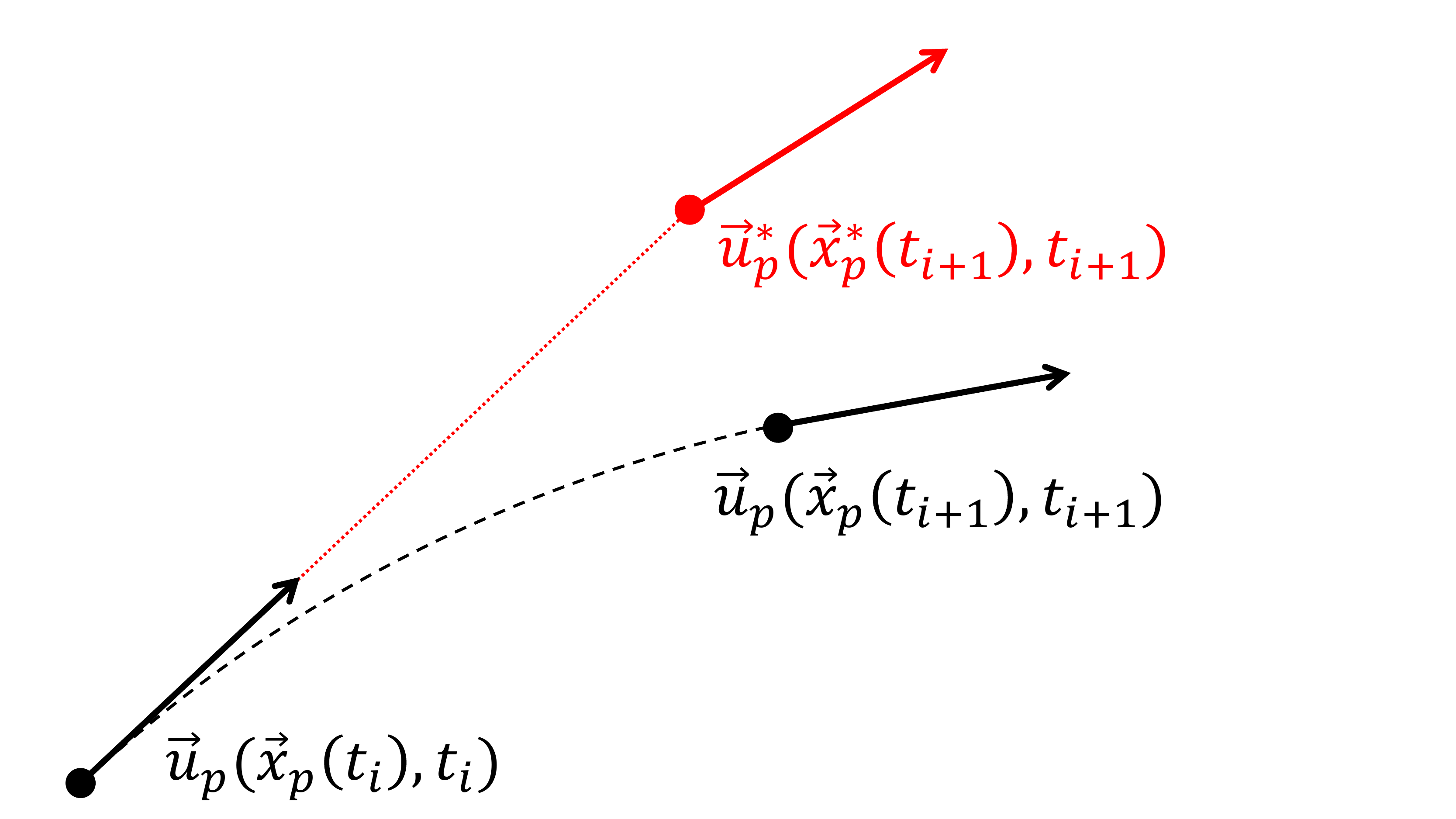}
\caption{Lagrangian scheme for the material acceleration calculation: true trajectory (in black) versus reconstructed trajectory (in red).}
\label{f:Lagrangian_path}
\end{figure}

In the \textit{queen2} algorithm, the material acceleration is evaluated based on a Lagrangian forward scheme (figure \ref{f:Lagrangian_path}), which consists in following a particle trajectory in a pseudo-tracking formulation:

\begin{equation}
\frac{D\vec{u}_p}{Dt}(\vec{x}_p(t_i),t_i) \simeq \frac{\vec{u}_p^{*}(\vec{x}_p^{*}(t_{i+1}),t_{i+1})-\vec{u}_p(\vec{x}_p(t_i),t_i)} {\Delta t}
\end{equation}

The particle velocity $\vec{u}_p^{*}(\vec{x}_p^{*}(t_{i+1}),t_{i+1})$  at a forward instant relies on a linear reconstruction of the particle path to determine its position $\vec{x}_p^{*}(t_{i+1})$, based on a trapezoidal scheme:

\begin{equation}
\vec{x}_p^{*}(t_{i+1}) \simeq \vec{x}_p(t_i)+ \bigg( \frac{\vec{u}_p(\vec{x}_p(t_i),t_i)+\vec{u}_p(\vec{x}_p(t_{i+1}),t_{i+1})}{2} \bigg)\Delta t
\end{equation}

In experiment 1, eight pairs of velocity fields covering a complete period are selected for the calculation of the instantaneous hydrodynamic stress tensor. As a second step, the stress tensor is projected on the fin surfaces, yielding a map of cartesian stress vectors which are then decomposed into biologically relevant components (along the fin principal axes - see figure \ref{f:fin}). The absolute values of the normal stress are averaged over these eight time frames to obtain a description of the period-averaged hydrodynamic load over the fin. In experiment 2, 32 pairs of velocity fields covering 6 cycles are used to evaluate the instantaneous stress distributions, and the period averaged absolute values of the normal stress are calculated based on these 32 time frames.

\subsection{Surface parametrization}
\label{sst:surface_param}

The raw images offer a very clear view of the foil midline curvature (see left panel of figure \ref{f:foil_kinematics} for example), and are used to parametrize the midline position with a polynomial of degree 2. For each instantaneous time frame under study, 25 points are manually superimposed on the fin, allowing to fit the quadratic function to the hydrofoil time-dependent deflection. The 3D hydrofoil, including the left and right trapezoidal sides plus the tip surface, is reconstructed based on the parametrized midline, assuming that deflection occurs only along one axis. This is a reasonable assumption since the foil length is about twice as large as its width.

A crucial step in our analysis is the projection of the total hydrodynamic stress tensor onto the different surfaces of the foil. The $i$-component of the stress (force per unit area) acting on a surface with a unit normal vector $\hat{n}$ (oriented outwards) is expressed as:

\begin{equation}
S_i(\vec{x},t,\hat{n})=s_{ij}(\vec{x},t) \cdot n_j
\end{equation}

The stress vector is initially expressed in cartesian coordinates. It is then decomposed into more relevant components, related to a set of orthogonal axes moving along with each fin surface, as illustrated in figure \ref{f:fin}. Anticipating useful analogies between our fin model and a real caudal fin, we employ the following terminology to define this coordinate system: the dorso-ventral axis $\hat{n}_{dv}$, the proximo-distal axis $\hat{n}_{pd}$ and the left-right axis $\hat{n}_{lr}$. These orthogonal sets are defined locally everywhere on the foil surfaces (sides and tip) and travel with it during the motion cycle. Thus, each grid point on the foil surfaces (25 nodes along the dorso-ventral axis, 25 nodes along the proximo-distal axis, 3 nodes along the left-right axis) is associated with a local set of perpendicular unit vectors ($\hat{n}_{pd}$, $\hat{n}_{lr}$, $\hat{n}_{dv}$), on which the stress vector is projected to give $S_{pd}, S_{lr}$ and $S_{dv}$ respectively.
%

In the present experimental cases, the fin operates in the inertial flow regime (although theoretically very close to the transitional range 300<$Re$<1000 \cite{McHenry2005}), where the normal stress component (dominated by the pressure) is typically larger than the viscous tangential stresses by at least two orders of magnitude. Depending on the flow under investigation, these distinct stress components may have different relative magnitudes. For example, the fish larvae evolve in a flow regime where the viscous shear stress is substantially larger \cite{McHenry2005}. Moreover, the tangential shear stress may have a particular significance in certain hydrodynamic problems, even when the flow is dominated by the fluid inertial forces. For instance, resolving the viscous shear stress maps on the surface of fin models might be informative in problems related to flow sensing in the skin of fish \cite{Konig2019}. In other words, despite the lower order of magnitude of the measured viscous tangential stress, these components may be of special interest for certain biological applications, which is why we retained them in our analysis.

The virtual object reconstructed from the parameterized deflection has a width larger than the real foil (2 mm instead of 0.55 mm) owing to the fact that the particles could not be resolved directly at the surface of the foil. It is necessary that the virtual surface remains inside the boundary layer of the real foil, in order to capture the hydrodynamic stresses correctly, in particular the shear stress which can vary significantly over a short distance inside that layer. The boundary layer thickness is defined as the distance, measured from the solid surface, at which the velocity reaches 99\% of the free-stream value \cite{Prandtl1952}:

\begin{equation}
\delta=y(u_x=0.99 u_\infty)
\end{equation}

The boundary layer of an oscillating fin can be associated to complex effects combining turbulence, separation and reattachment \cite{Obremski1967, Kobashi1980, Arnal1984, Incropera2007, Kunze2011}. Nevertheless, these flow phenomena would tend to extend the boundary layer over a larger region above the solid surface, therefore, we use the thickness of a laminar boundary layer as a conservative assumption to ensure that the reconstructed foil boundary is sufficiently close to the real surface where we want to extract the stresses. The solution for a laminar boundary layer over a flat plate is \cite{Prandtl1952}:

\begin{equation}
\delta=5 \sqrt{\frac{\nu x}{U}}
\label{eq:BL}
\end{equation}

The value of $x$ is the distance along the proximo-distal axis measured from the foil leading edge. The stress maps and curves presented in section \ref{sst:results_stress} correspond to locations $x$ on the fin in the range of [0.25$L$, $L$], where $L$ is the total length of the foil. Even for the thinnest boundary layer, which should occur in exp. 2 where the average velocity magnitude is the largest, the boundary layer thickness evaluated by equation \ref{eq:BL} remains in the range of [2.3, 4.6] mm. Hence, the hydrodynamic stress values extracted at 0.725 mm away from the real surface are computed well inside the boundary layer and should provide an accurate representation of the real surface distribution. Although the magnitudes might be slightly shifted, the stress patterns along the different axes of the fin are expected to remain accurate.

\subsection{Euler-Bernoulli beam theory}
\label{sst:euler_bernoulli}

A complementary approach is employed in exp. 1 to derive the normal force distribution on the synthetic fin, in order to provide a validation for the PTV-based hydrodynamic stresses. Since the deflections remain small ($\theta_0$=10$\degree$ at the peduncle), it can be assumed that the hydrofoil obeys a linearized Euler-Bernoulli equation of motion, where the force per unit length normal to the surface ($f_n$) is given by \cite{Paraz2016}:

\begin{equation}
\begin{split}
f_n(x,t)=m(x) \frac{\partial^2 h}{\partial t^2} +  \frac{\partial^2}{\partial x^2} \Big(  E I(x) \frac{\partial^2 h}{\partial x^2}  \Big) +\alpha \frac{\partial h}{\partial t} \\
+ \eta  \frac{\partial}{\partial t}\Big( \frac{\partial^4 h}{\partial x^4}  \Big)
\label{eq:euler_bernoulli}
\end{split}
\end{equation}

Despite the fact that a quadratic function offers a good fit to the deflection profiles as described above, for a comparison with Euler-Bernoulli theory with stress-profiles another function $h(x,t)$ has to be used. This is because a quadratic function would lack a non-zero fourth order spatial derivative and hence would not show a force profile. Therefore the time-dependent deflections of the hydrofoil were once more fitted based on the raw images, but this time, with a function inspired by the general solution for a freely vibrating beam, clamped at one end, which is given by:

\begin{equation}
\begin{split}
h(x,t) = C_1 sin(\frac{2\pi t}{T}+ \phi_1)sinh(\beta x) \\
+ C_2 sin(\frac{2\pi t}{T}+ \phi_2)sin(\beta x)
\label{eq:deflection_fit}
\end{split}
\end{equation}

The first term on the right hand side of equation \ref{eq:euler_bernoulli} is the inertial force, the second term, the elastic restoring force, the third term, the linear fluid damping and the fourth term, the internal viscoelastic damping, with $m(x)$, the foil density per unit length, $E$, Young's modulus, $I(x)$, the second moment of area, $\alpha$ and $\eta$, the fluid and internal viscous damping coefficients per unit length, respectively. The foil thickness $d$, its width $B(x)$ and its density $\rho_f$ are used to evaluate $m(x)$ and $I(x)$:

\begin{equation}
I(x)=\frac{B(x) d^3 }{12} \hspace{2cm} m(x)=\rho_f B(x) d
\end{equation}

The boundary condition at $x=0$, where the pitching motion $\theta(t)$ is imposed, can be expressed as:

\begin{equation}
h(x=0,t)=0  \hspace{2cm} \frac{\partial h}{\partial t}(x=0,t)=\theta(t)
\end{equation}

The boundary conditions at the free tip are not explicitly taken into account, since rather than solving equation \ref{eq:euler_bernoulli} for $h(x,t)$, we use the known deflection to calculate the time-dependent load on the fin.

The fluid damping coefficient is determined based on the decay exponent $\gamma$ obtained from the impulse response tests (equation \ref{eq:damping}):

\begin{equation}
\alpha=\gamma \rho_f  B(x) d
\end{equation}

This yields values of 0.009 kg m\textsuperscript{-1}s\textsuperscript{-1} in air and 0.03 kg m\textsuperscript{-1}s\textsuperscript{-1} in water, with a ratio of roughly 3.3 between both. An approximate ratio of 20 was found between $\alpha_{water}$ and $\alpha_{air}$ in a previous study, using a similar set up and a polysiloxane foil (with comparable thickness but larger area and larger Young's modulus) \cite{Paraz2016}. Based on a modal analysis of the beam deflection, a mathematical relation can be established between $\eta$, $\alpha$, $\gamma$ and $\omega_0$ (see equations B3 and B4 in \cite{Paraz2016}). Inserting our measured values of $\omega_0$ and $\gamma$ (in air) into that equation, we estimated an internal damping coefficient ($\eta$) for our PDMS fin of the order of 10\textsuperscript{-10} Nm\textsuperscript{2}s. The internal damping contribution in equation \ref{eq:euler_bernoulli} is thus expected to be negligible compared to the other terms.

\subsection{Control volume analysis}
\label{sst:control_volume}

A control volume analysis is applied to measure the forces acting on the fluid affected by the fin motion \cite{Koochesfahani1989, Anderson1998, Dabiri2005, Godoy-Diana2008, Hong2008, Bohl2009, Shinde2014} and validate the consistency of the stress distributions obtained on the fin surface. The total force exerted by the fin on the fluid can be expressed as the rate of change of fluid momentum inside the control volume, summed with the hydrodynamic stress tensor projected on the surface of that domain \cite{Noca1996, Unal1997, Batchelor2000}:

\begin{equation}
\vec{F}= \iiint_V \frac{\partial (\rho \vec{u})}{\partial t} dV + \oiint_S (\hat{n} \cdot \vec{u}) \rho \vec{u} dA  - \oiint_S \hat{n} \cdot \bm{s} dA
\label{eq:control_volume}
\end{equation}

We used this approach as a complementary calculation to derive the time-dependent total force and verify the coherence of the PTV-based stress data. The control volume was defined as the whole 3D measurement domain, except the last two grid planes on each outer boundary walls. The dimensions of the control volume are restricted by the field of view of the cameras, which limits the inclusion of the whole body of fluid affected by the motion of the hydrofoil, especially in the streamwise direction. Therefore, we would expect a certain offset of the streamwise force ($F_x$) compared to the more accurate values resulting from the integration of the stress distributions directly on the hydrofoil surfaces.

\subsection{Uncertainty analysis}
\label{sst:uncertainty_calculation}

A central point of our work is the evaluation of uncertainties on the hydrodynamic stresses, to ensure that stress maps can be measured accurately on the surface of a small synthetic fin with biologically relevant kinematics. To determine the positional uncertainty arising from the triplet reconstruction scheme, we consider the coordinates of a particle match, where the physical coordinates $X_i$ of the dewarped particles images are averaged over the different cameras $N_{cam}$. From the uncertainty of each $X_i$ given by several parameters but dominated by the least square errors ($LSE$) from the calibration and the 2D Gaussian fit during the particle identification step, we can then obtain an uncertainty in position $\sigma_x$ using error propagation given by: $\sigma_x=\sigma_y=\frac{LSE}{\sqrt{N_{cam}}}\simeq 3.6 \mu \textrm{m}$

The uncertainty in the $z$-direction ($\sigma_z$) is computed using the reference plane distance from the face plate of the cameras mount ($\sim$475 mm) and the distance between the apertures for a half-angle of about 6.5\degree, leading to $\sigma_z \simeq$ 32 $\mu$m. The positional uncertainty propagates to the velocity components:

\begin{equation}
\sigma _{u_i}=\frac{\sqrt{2}\sigma_i}{\delta t}
\end{equation}

The time delay $\delta t$ between two laser pulses is considered as exact, resulting in $\sigma_{u_x}=\sigma_{u_y} \simeq$ 0.002 m/s and $\sigma_{u_z} \simeq$ 0.018 m/s. During the velocity interpolation onto a regular 3D grid, appropriate filters were applied (see section \ref{sst:PTV}). Therefore, these uncertainties are a conservative overestimation.

Noise propagation from the velocity field to the material acceleration and to the integrated pressure field has been the object of many studies \cite{Liu2006, Violato2011, vanOudheusden2013, deKat2013}. Based on statistical arguments, a normal probability distribution can be determined for the pressure uncertainty, assuming a Lagrangian material acceleration, a normal probability distribution for the velocity uncertainty $\mathcal{N}(\mu_u=0,\sigma_u^2)$ and integration paths covering $n$ nodes \cite{Wang2017}:

\begin{equation}
\mathcal{N}(n \rho \mu_u \Delta x \Delta t^{-1}, n \rho^{2} \sigma_u^2 \Delta x^2 \Delta t^{-2}/2)
\end{equation}

Accordingly, the pressure error in the case of a zero-expectation value ($\mu_u$=0) should not exceed the standard deviation:

\begin{equation}
\sigma_p \leqslant \sqrt{\frac{n}{2}}\frac{\rho \sigma_u \Delta x}{\Delta t}
\label{eq:error_pressure_statistical}
\end{equation}

The Eulerian method would produce less clear error distributions due to the nonlinear advective term of the material acceleration in equation \ref{eq:material_derivative}, making it hard to define the uncertainty without \textit{a priori} knowledge of the flow velocity field \cite{Wang2017}.

The main sources of uncertainty on the Lagrangian material acceleration are the truncation error associated to the numerical discretization, the precision error propagated from the velocity data and the pseudo-tracking scheme inaccuracy \cite{deKat2012, vanOudheusden2013, Laskari2016}. The discrepancy between the true and estimated particle positions $\vec{x}_p$ and $\vec{x}_p^{*}$ in the Lagrangian pseudo-tracking approach results in a discrepancy between the real and approximated velocities $\vec{u}_p$ and $\vec{u}_p^{*}$ which adds up to the measurement uncertainty:

\begin{equation}
\sigma_{u_i^{*}}^2=\sigma_{u_i}^2+ \big( (\vec{u}_p^{*}-\vec{u}_p)\cdot \hat{x}_i \big)^2
\label{eq:total_error_velocity}
\end{equation}

\begin{equation}
\vec{u}_p^{*}-\vec{u}_p \simeq \big( (\vec{x}_p^{*}-\vec{x}_p)\cdot \vec{\nabla} \big)\vec{u}
\label{eq:error_Lagrangian_velocity}
\end{equation}

Equations \ref{eq:total_error_velocity} and \ref{eq:error_Lagrangian_velocity} can be simplified by using the following approximation:

\begin{equation}
(\vec{x}_p^{*}-\vec{x}_p)\cdot \hat{x}_i \simeq \sigma_{u_i} \Delta t
\end{equation}

\begin{equation}
\begin{split}
\implies  (\vec{u}_p^{*}-\vec{u}_p)\cdot \hat{x}_i = \Delta t (\vec{\sigma}_u \cdot \vec{\nabla})u_i \\
\mathrm{with\:} \vec{\sigma}_u=(\sigma_{u_x},\sigma_{u_y},\sigma_{u_z})
\end{split}
\end{equation}

A forward difference scheme is used to evaluate the material acceleration based on $\vec{u}_p(t)$ and $\vec{u}_p^{*}(t+\Delta t)$, yielding a total uncertainty on the material acceleration (with $a_i=Du_i/Dt$):

\begin{equation}
\sigma_{a_i}^2 = \frac{\sigma_{u_i}^2+\sigma_{u_i^{*}}^2}{\Delta t^2}
= \frac{2\sigma_{u_i}^2}{\Delta t^2} + \big( (\vec{\sigma}_u \cdot \vec{\nabla})u_i \big)^2
\label{eq:error_material_acceleration}
\end{equation}

It contains the error propagated from the velocity field and a second contribution from the path line reconstruction inaccuracy, whereas the truncation error from the interpolation of the velocity data, associated to the spatial resolution, was considered negligible compared to the other terms.

The uncertainty on the pressure is proportional to the uncertainty on the material acceleration, the spatial resolution $\Delta x_i$ and the number of nodes $n$ crossed along the integration path. Moreover, the algorithm involves a median polling among a collection of eight paths. Since it is impossible to predict which direction will correspond to the selected median, averaging the uncertainty in all three directions is a reasonable approach. Besides, given a variable $X$ with a statistical sample of size $N$ and a median value $\widetilde{X}$, a mathematical relation between the variances is obtained \cite{Kenney1962}:

\begin{equation}
\sigma_{\widetilde{X}} = \sqrt{\frac{\pi}{2(N-1)}} \sigma_X
\end{equation}

This yields an approximation for the pressure uncertainty (with the indices 1,2,3 referring to the three cartesian directions):

\begin{equation}
\implies \sigma_p = \frac{\rho}{3} \sqrt{\frac{\pi}{14}}\sum_{i=1}^{3}n \Delta x_i \sqrt{\frac{2\sigma_{u_i}^2}{\Delta t ^2}
+\big( (\vec{\sigma}_u \cdot \vec{\nabla})u_i \big)^2}
\label{eq:error_pressure}
\end{equation}

The path length $n \cdot \Delta x_i$ typically covers half of the domain size, which is of the order of 20 mm in the $x$-$y$ plane, and 8 mm in the $y$-$z$ plane. The spatial dependence of pressure errors is analyzed in section \ref{sst:results_uncertainty}.

The discrete formulation of the viscous shear stress (equation \ref{eq:discrete_shear_stress}) is associated to an uncertainty:

\begin{equation}
\sigma_{\tau_{ij}} \simeq \frac{\mu}{\Delta x} \sqrt{\frac{\big( \sigma_{u_i}^2 + \sigma_{u_j}^2 \big)}{2}}
\label{eq:error_shear_stress}
\end{equation}

Finally, the uncertainties are reduced by an additional factor of $\sqrt{N_{time} \times N_{spatial}}$, where $N_{time}$ is the number of time frames involved in the period-averaging step and $N_{spatial}$ is the number of grid points used in the spatial average. For the instantaneous stress curves of exp. 1, spatial averaging is performed over the most distal portion (15\%) of the fin sides, covering the last four rows of points ($N_{spatial}=4$), whereas the stress curves extracted on the tip surface are obtained by averaging over the three rows of points covering that surface ($N_{spatial}=3$). In the case of the period-averaged distributions, $N_{spatial}=4$, as each stress curve is averaged over the left and right sides and over the dorsal and ventral portions of the foil.

\subsection{Temporal and spatial resolutions}
\label{sst:resolution}

The time step between subsequent particles images ($\delta t$) is limited on one side by the error propagation to the velocity field and on the other side, by the characteristic time scale of the flow (particles must remain traceable). The value of 2.5 ms was found to provide a good trade-off between both constraints. A similar twofold effect is associated with $\Delta t$ and $\Delta x$, which must remain small enough to permit the Taylor expansions inherent to the calculation of the viscous shear stress and Lagrangian acceleration, but not too small to cause unreasonably large errors propagated from the velocity field. Additionally, the reliability of the pseudo-tracking approach underlying the Lagrangian acceleration can become an issue in the case of strongly curved trajectories involved in highly rotational fluid \cite{Liu2006, vanOudheusden2013}. The simple criterion $\Delta t \geqslant \Delta x/U_{ref}$ was suggested, where $U_{ref}$ is the characteristic velocity of a flow structure \cite{Wang2017}. In our experiments, the characteristic flow structures are vortices generated by the flapping foil. An estimate of $U_{ref}$ is therefore obtained by the examination of 2D slices from instantaneous velocity fields, of which an example is illustrated in the left panel of figure \ref{f:velocity_color_map_2D}. The velocity magnitude of the characteristic flow structures is of the order of 0.075 m/s, implying a ratio $\Delta x /U_{ref}$ of about 0.01 s for exp. 1 (it is even smaller for exp. 2 where the vortical motion is stronger). Therefore, the condition stated above is met with a temporal resolution of $\Delta t$=0.0125 s. Another criterion for vortex dominated flows was proposed, stating that a vortex should not execute more than half a turn during $\Delta t$, with a recommended a ratio of acquisition frequency ($f_{acq}=1/\Delta t$) to turnover frequency of at least 10\cite{deKat2012}. In our study, the acquisition frequency is 80 s\textsuperscript{-1} and the turnover frequency can be defined as $\abs{\vec{u}}_{ave}/a$, yielding values of 1.4 s\textsuperscript{-1} and 2.07 s\textsuperscript{-1} for exp. 1 and 2, respectively, which shows that the criterion is fulfilled.

For the \textit{queen2} algorithm, pressure errors in the range of 5-10\% were reported for $\Delta x <$0.0625$D$, where $D$ is the characteristic object dimension \cite{Dabiri2014}. The flapping foil in our experiments covers a displacement of 11-14 mm, which translates into a ratio of $\Delta x/D$ between 0.054 and 0.068, indicating that the grid spacing is small enough to produce reasonably small pressure errors. Lastly, a recommendation was made to ensure the validity of the null-pressure boundary condition inherent to the \textit{queen2} algorithm, based on the condition $H/D \geqslant 2$ where $H$ is half of the domain size \cite{Dabiri2014}. Substituting the tip excursion amplitudes (11-14 mm) for the characteristic dimension $D$ and replacing $H$ by the average half domain size in the $x$ and $y$ directions, we conclude that the size of the measurement domain lies right above that limit.

\section{Results}

\subsection{Instantaneous and period-averaged stress distributions}
\label{sst:results_stress}

\subsubsection{Experiment 1}

\begin{figure}
\centering
\includegraphics[width=3cm, trim={0.5cm -3cm 0.5cm 3cm}]{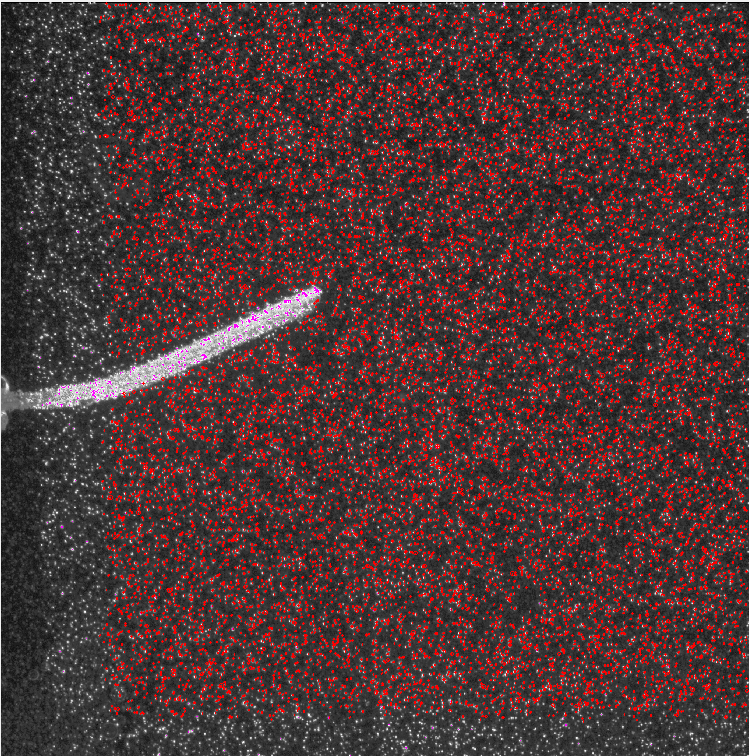}
\includegraphics[width=5cm, trim={0cm 0cm 1.5cm 0cm}]{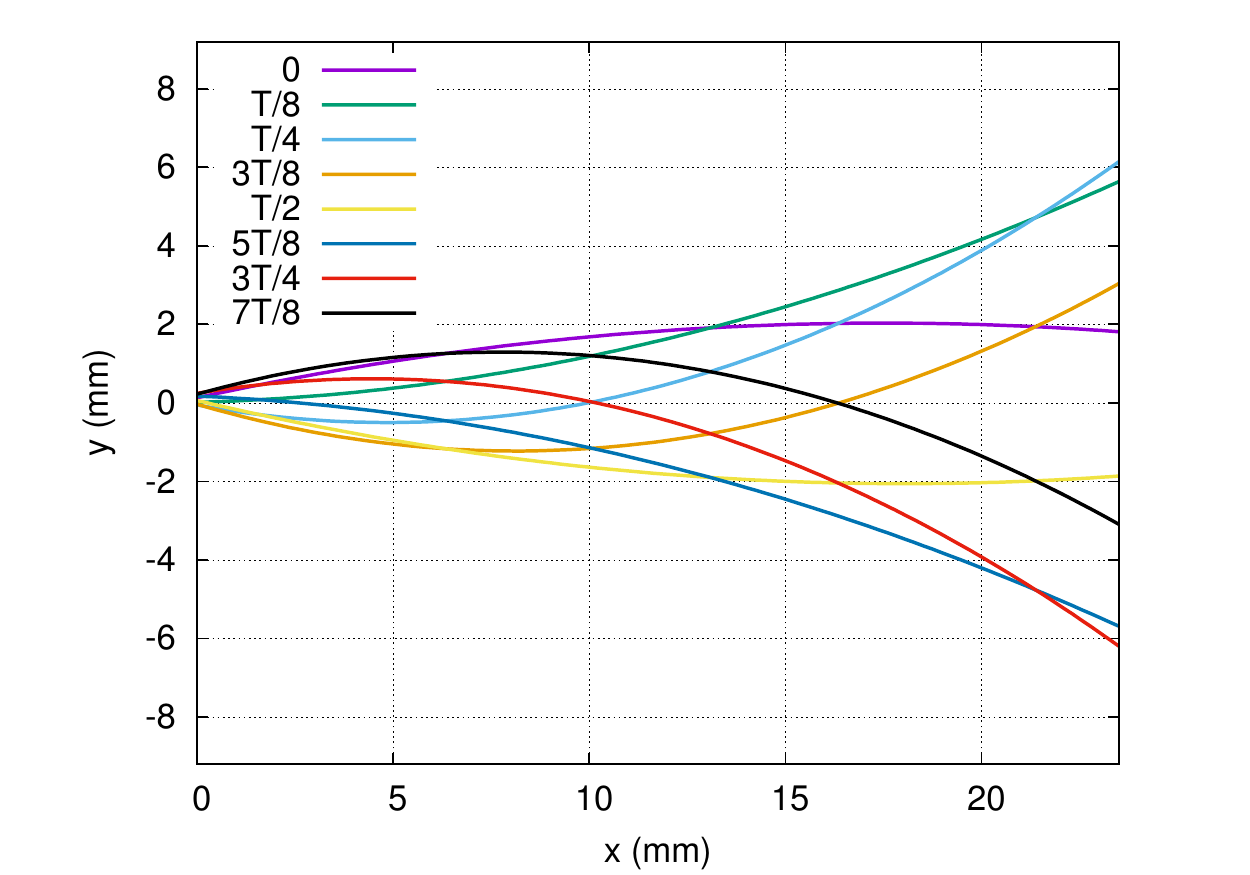}
\caption{\textit{Left}: 3D reconstructed particles positions (red points) projected on a single raw image from the triplet. \textit{Right}: Foil midline positions in exp. 1, fitted with a polynomial of degree 2.}
\label{f:foil_kinematics}
\end{figure}

The parametrized fin deflections are presented in figure \ref{f:foil_kinematics} (right panel), at eight selected time points for which the stress distributions are presented. The most proximal portion of the fin is excluded from the 3D particles positions field (figure \ref{f:foil_kinematics}, left panel), and in the following graphs, the zero of the $x$ axis is relocated at the initial point on the fin where the flow quantities are extracted.

\begin{figure}
\centering
\includegraphics[width=8cm, trim={2cm 0.5cm 0.5cm 0.5cm}]{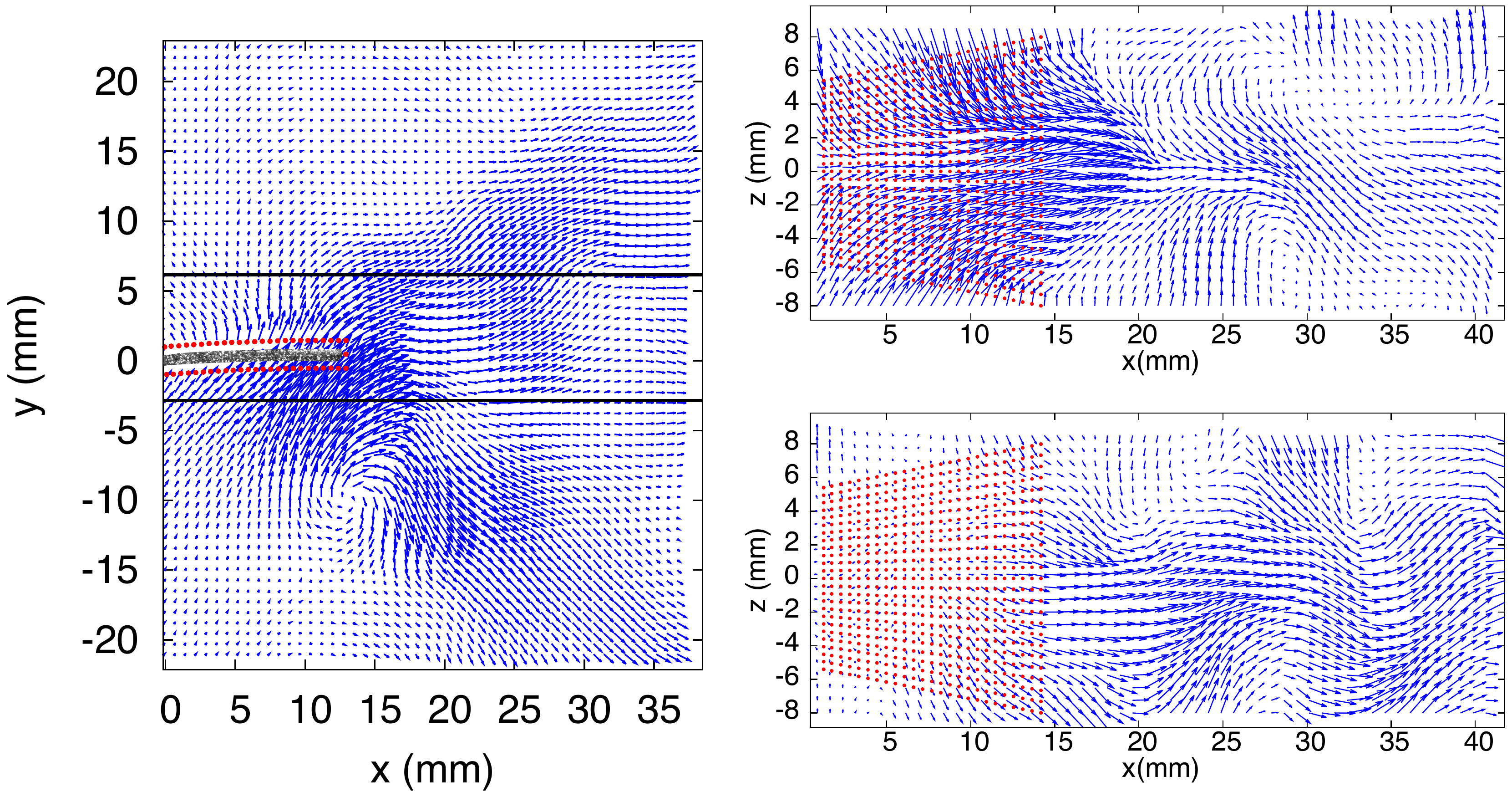}
\caption{Velocity vectors field - exp. 1, $t_0$. \textit{Left panel}: Vertical midplane, with image of the real fin superimposed inside the virtual boundary (red points). \textit{Right panel}: Horizontal planes indicated by black lines in the left panel.}
\label{f:velocity_vectors}
\end{figure}

\begin{figure}
\centering
\includegraphics[width=8cm, trim={2cm 0.5cm 0.5cm 0.5cm}]{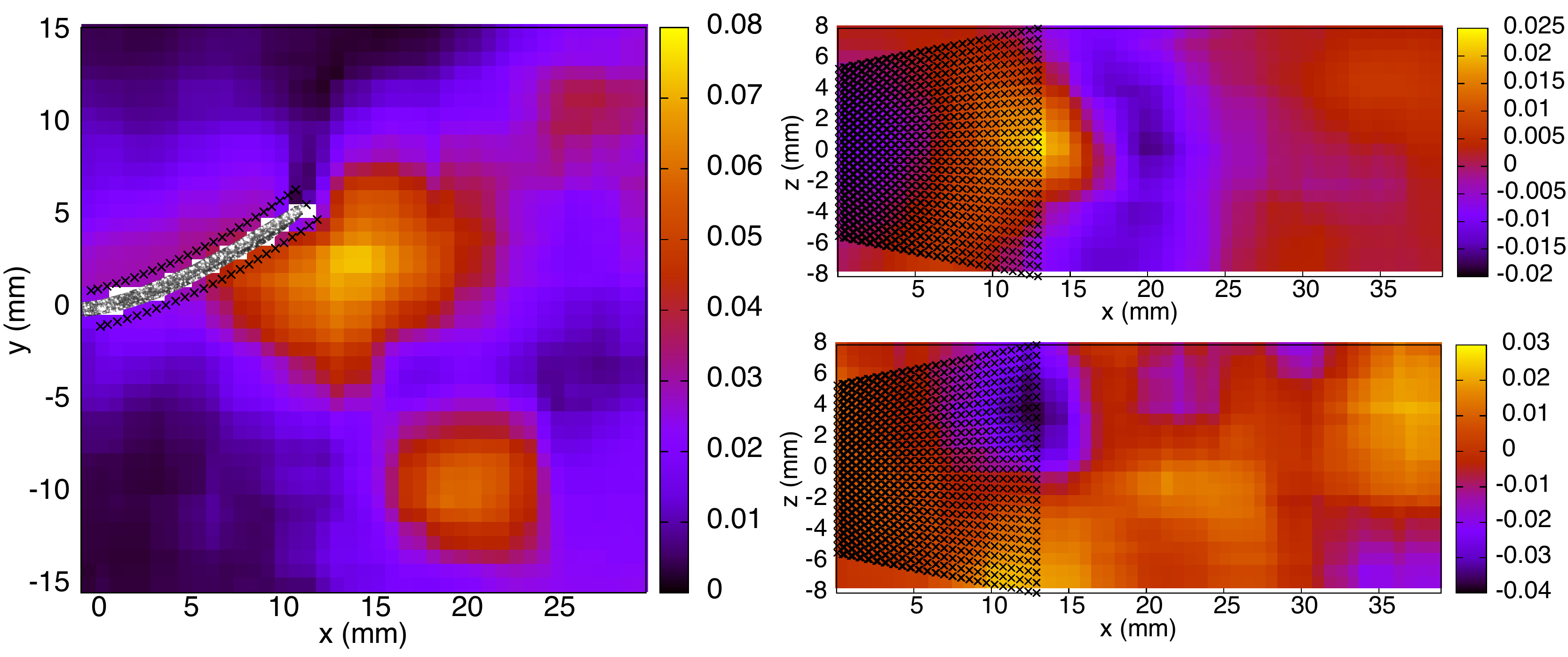}
\caption{Velocity color maps (m/s) - exp. 1, $t_{1/4}$. \textit{Left panel}: Velocity magnitude in the vertical midplane. \textit{Right, top and bottom panels}: $y$ and $z$ velocity components, respectively, in a horizontal plane below the foil ($y$=-5.275 mm).}
\label{f:velocity_color_map_2D}
\end{figure}

The velocity field at $t_0$ is shown in figure \ref{f:velocity_vectors}, where 2D slices from the measurement volume are presented: one vertical plane extracted in the middle of the domain, and two horizontal planes located slightly below and above the fin. In both vertical and horizontal planes, the shedding of counterrotating vortices can be observed, corresponding to the 2D projections of vortex rings. In figure \ref{f:velocity_color_map_2D}, the velocity field at $t_{1/4}$ is represented by a color map. The velocity magnitude is depicted in the vertical midplane (left panel), where high values are associated to two vortices, one previously shed in the downstream wake and a second still in formation, attached to the foil. The vertical ($u_y$) and transversal ($u_z$) velocity components in a single horizontal plane are shown in the right panel of figure \ref{f:velocity_color_map_2D}.

\begin{figure}
\centering
\includegraphics[width=5.5cm]{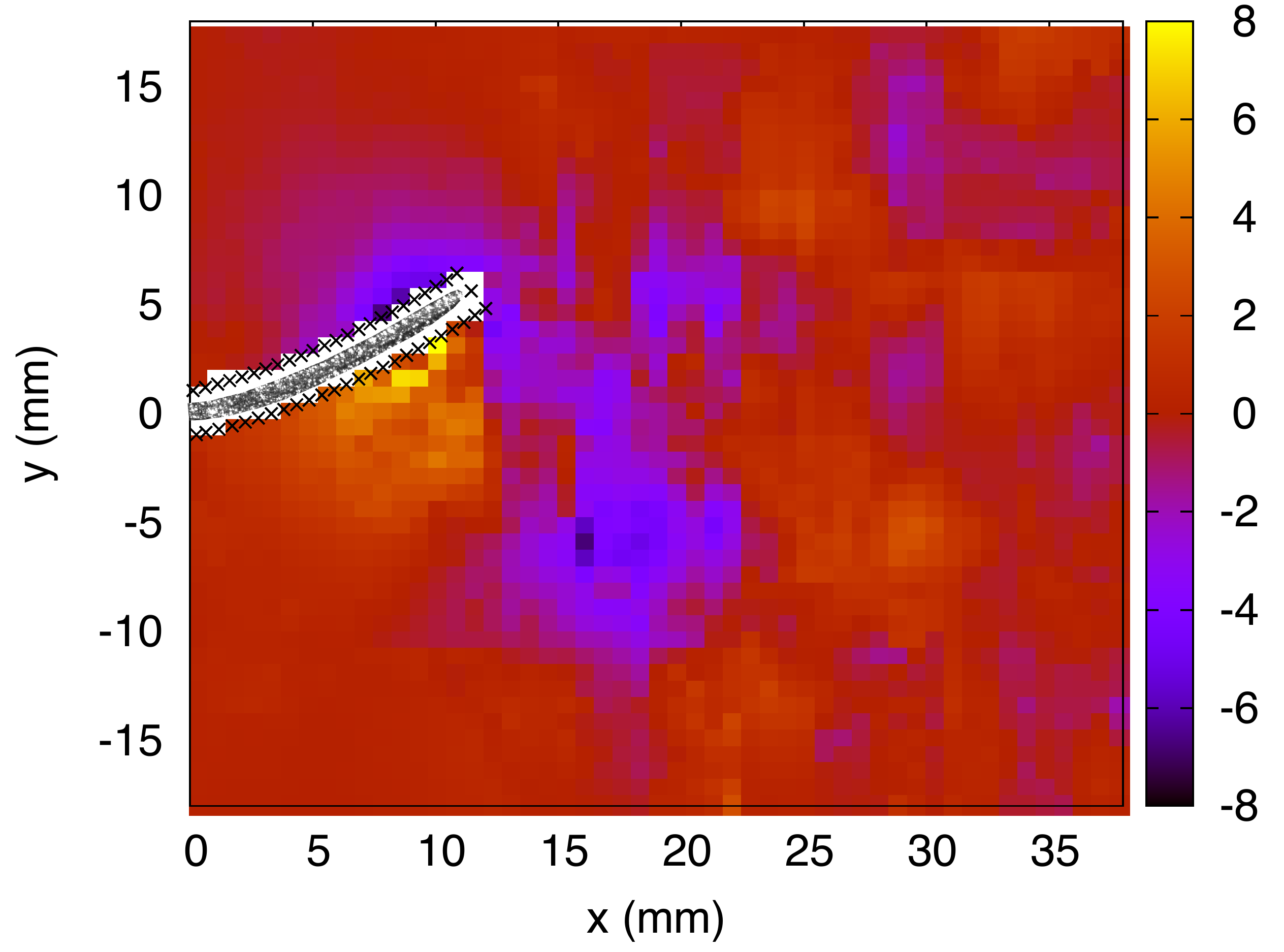}
\includegraphics[width=5.5cm]{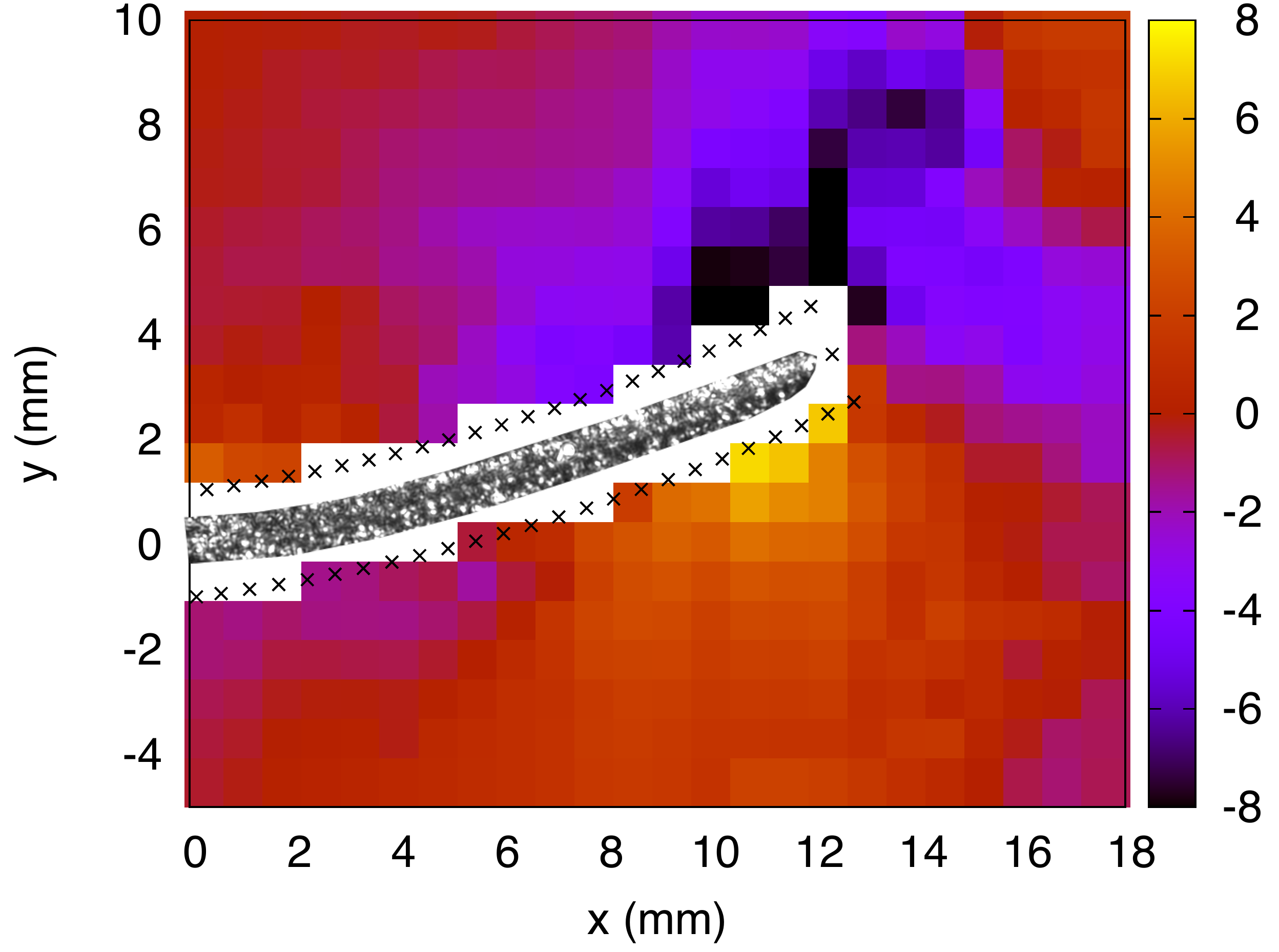}
\caption{Pressure color map (Pa) in the vertical midplane (exp. 1). \textit{Top panel}:  $t_{1/4}$. \textit{Bottom panel}: $t_{3/8}$ (magnified region around the fin).}
\label{f:t1_4_t3_8_2D_pressure}
\end{figure}

\begin{figure}
\centering
\includegraphics[width=5.5cm, trim={0.5cm 1cm 0.5cm 1cm}]{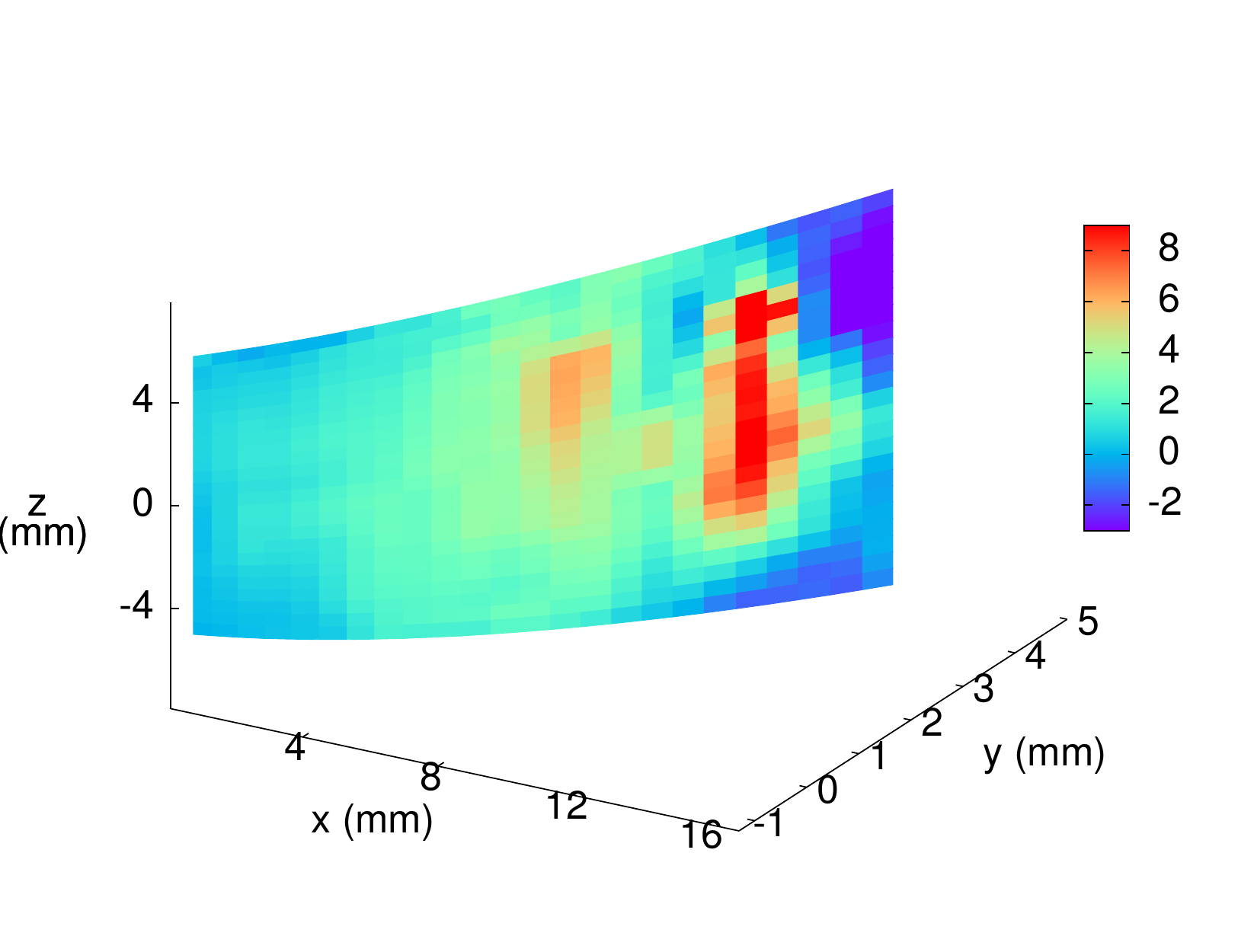}
\includegraphics[width=5.5cm, trim={0.5cm 1cm 0.5cm 1cm}]{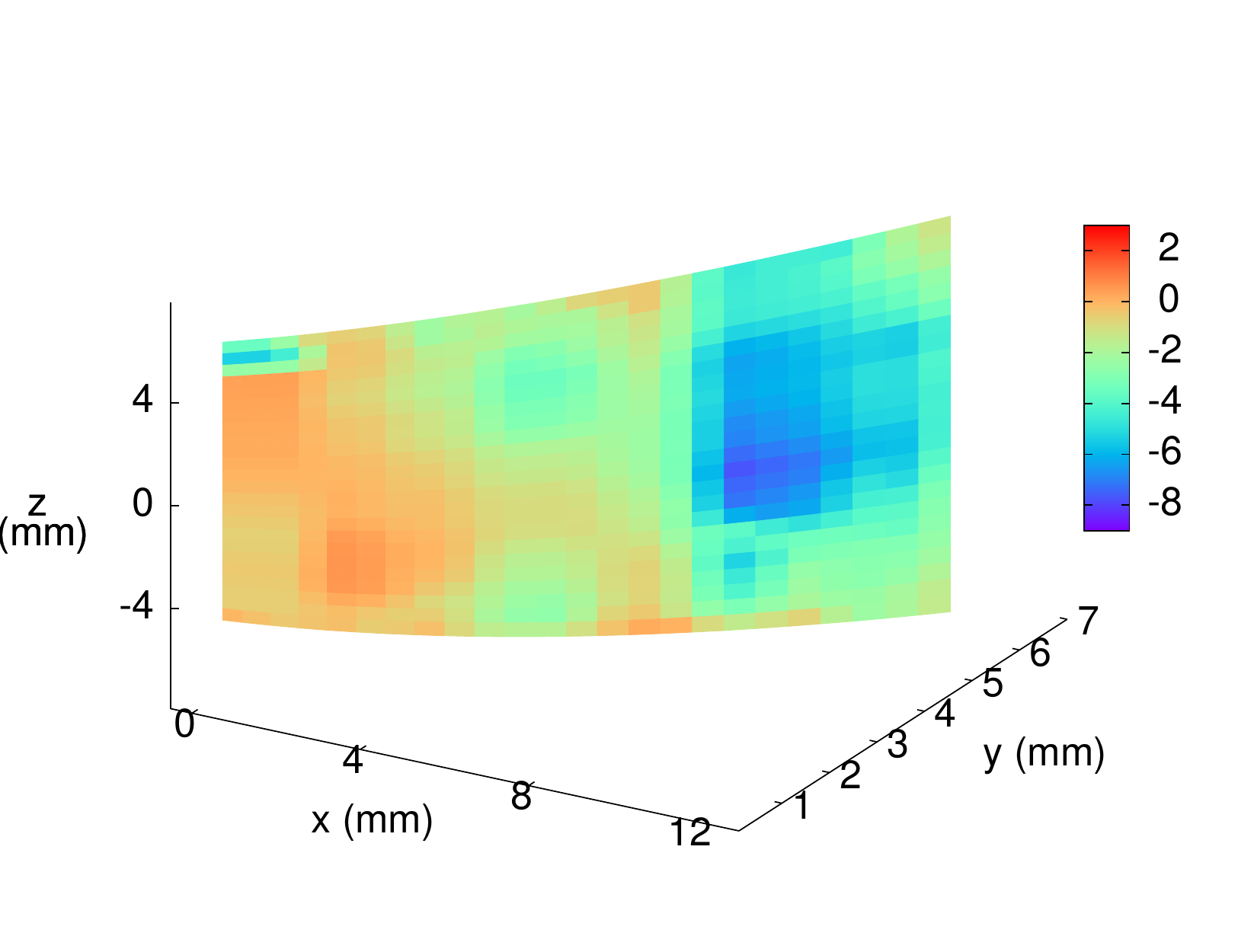}
\caption{Pressure color maps (Pa) on the left and right sides surfaces of the fin, respectively in top and bottom panels - exp. 1, $t_{1/4}$}
\label{f:t1_4_surface_pressure}
\end{figure}

\begin{figure}
\centering
\includegraphics[width=5.5cm, trim={0.5cm 1cm 0.5cm 1cm}]{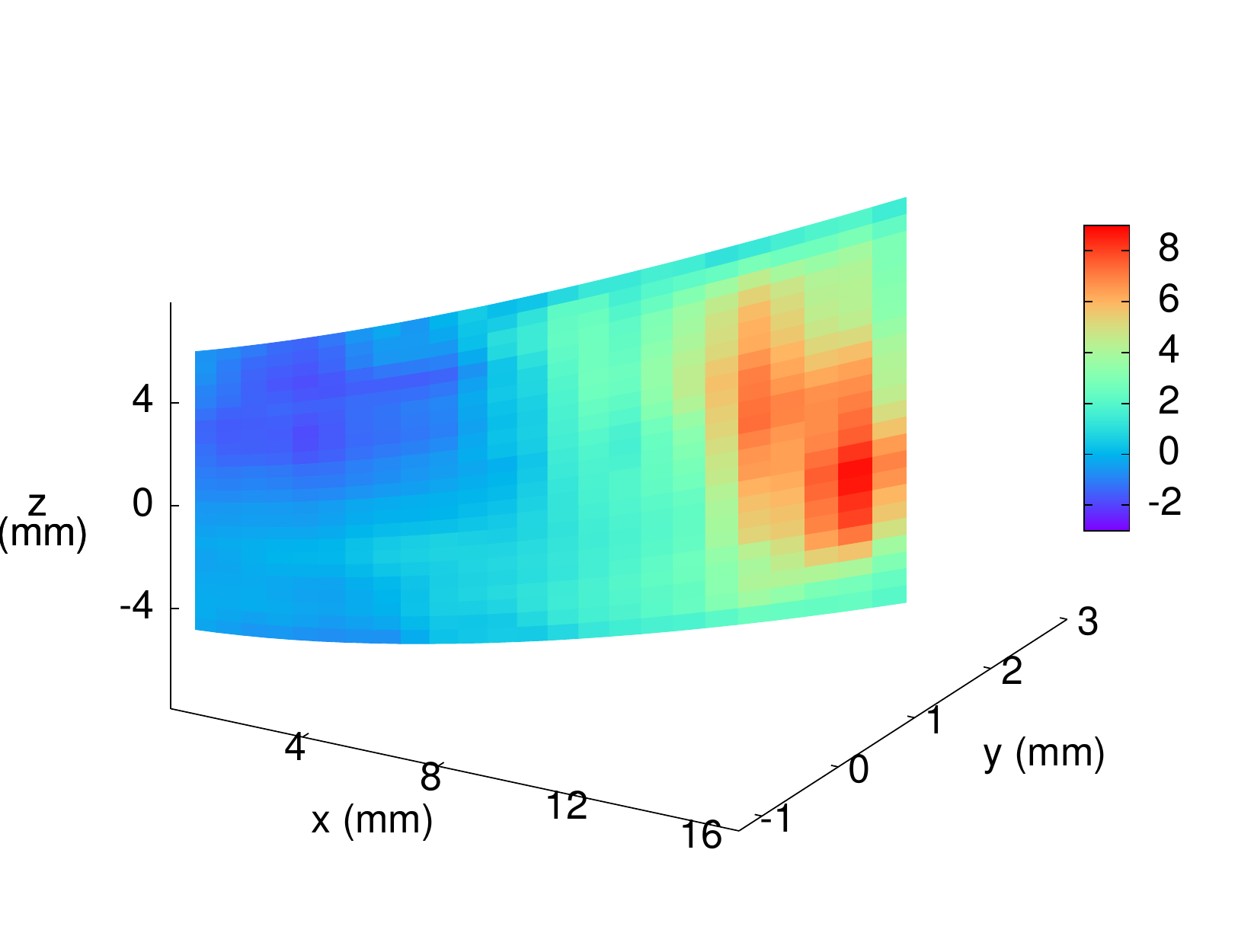}
\includegraphics[width=5.5cm, trim={0.5cm 1cm 0.5cm 1cm}]{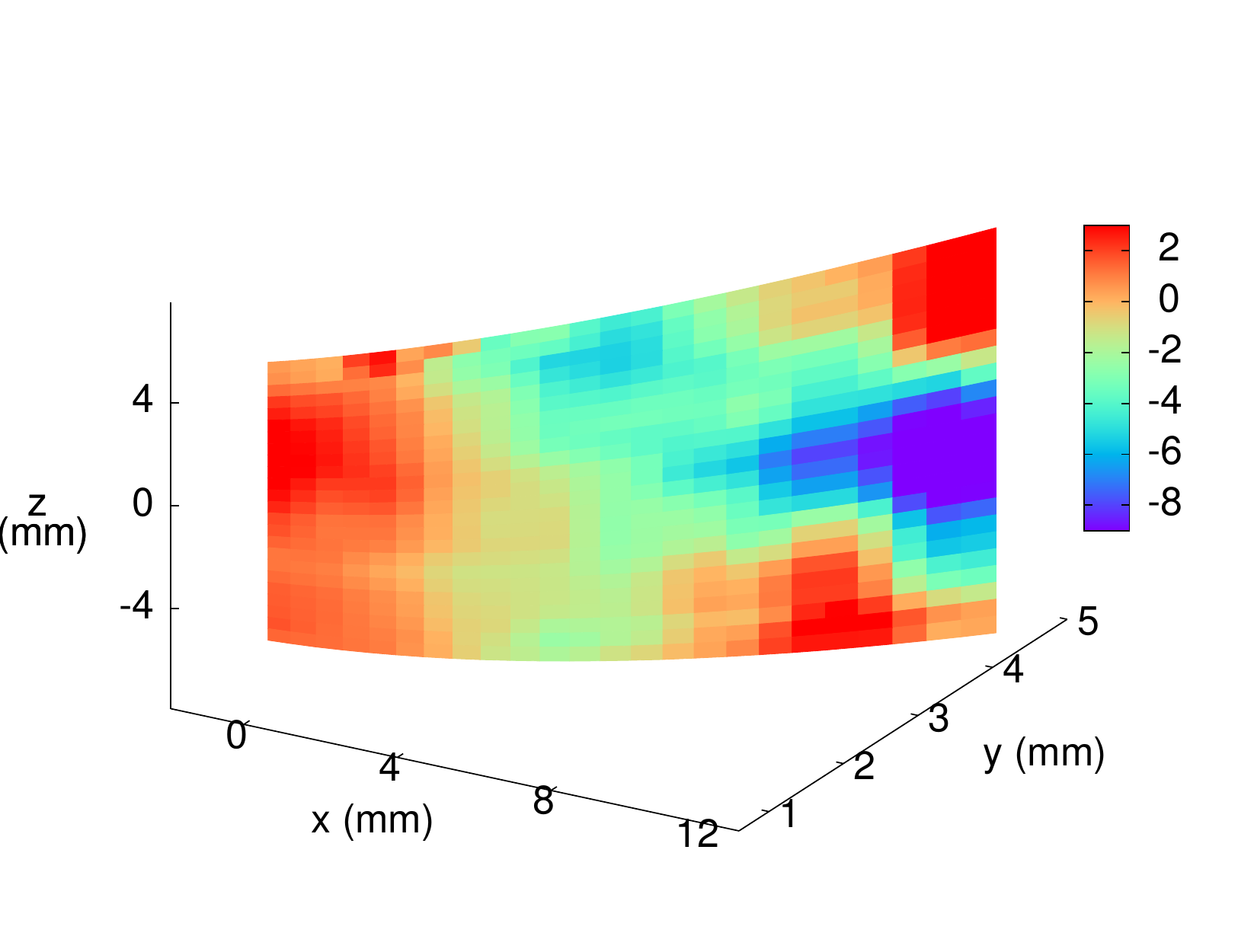}
\caption{Pressure color maps (Pa) on the left and right sides surfaces of the fin, respectively in top and bottom panels - exp. 1, $t_{3/8}$}
\label{f:t3_8_surface_pressure}
\end{figure}

A pair of subsequent time points ($t_{1/4}$ and $t_{3/8}$) is analyzed to ensure that the stress profiles on the fin surface present a realistic time-evolution (no unphysical abrupt transitions). Figure \ref{f:t1_4_t3_8_2D_pressure} offers an overview of the pressure distributions in the vertical midplane extracted from the 3D domain. The fin is traveling towards its left side (negative $y$ values). A vortex is visible close to the location ($x=17.5$mm, $y=-7.5$mm) in figure \ref{f:t1_4_t3_8_2D_pressure} (top panel), associated with a core of negative pressure. We are especially interested in the stress maps directly on the fin surfaces. Figures \ref{f:t1_4_surface_pressure} and \ref{f:t3_8_surface_pressure} present the pressure distributions on the left and right sides of the foil (at $t_{1/4}$ and $t_{3/8}$). As can be deduced from figure \ref{f:t1_4_surface_pressure}, the left side of the foil at $t_{1/4}$ is subjected to positive pressure, yielding a normal stress directed towards its surface. The maximum value is of the order of 10 Pa, located distally on the fin, although not at the very end, and centrally along the dorso-ventral axis. The tip corners experience negative pressure with values around -4 Pa. A similar trend is observed for the left side in the following time frame ($t_{3/8}$), where the pressure maximum is reduced to a value around 8 Pa (figure \ref{f:t3_8_surface_pressure}, top panel). At $t_{1/4}$, the right side of the fin is subjected to negative pressure on its most distal portion, of the order of -7 Pa (figure \ref{f:t1_4_surface_pressure}, bottom panel), an effect that concentrates more centrally along the dorso-ventral axis at $t_{3/8}$  (figure \ref{f:t3_8_surface_pressure}, bottom panel), with a minimum decreasing to approximately -15 Pa, along with the appearance of two small positive pressure spots of about 5 Pa close to the dorsal and ventral corners.

\begin{figure}
\centering
\includegraphics[width=5.5cm]{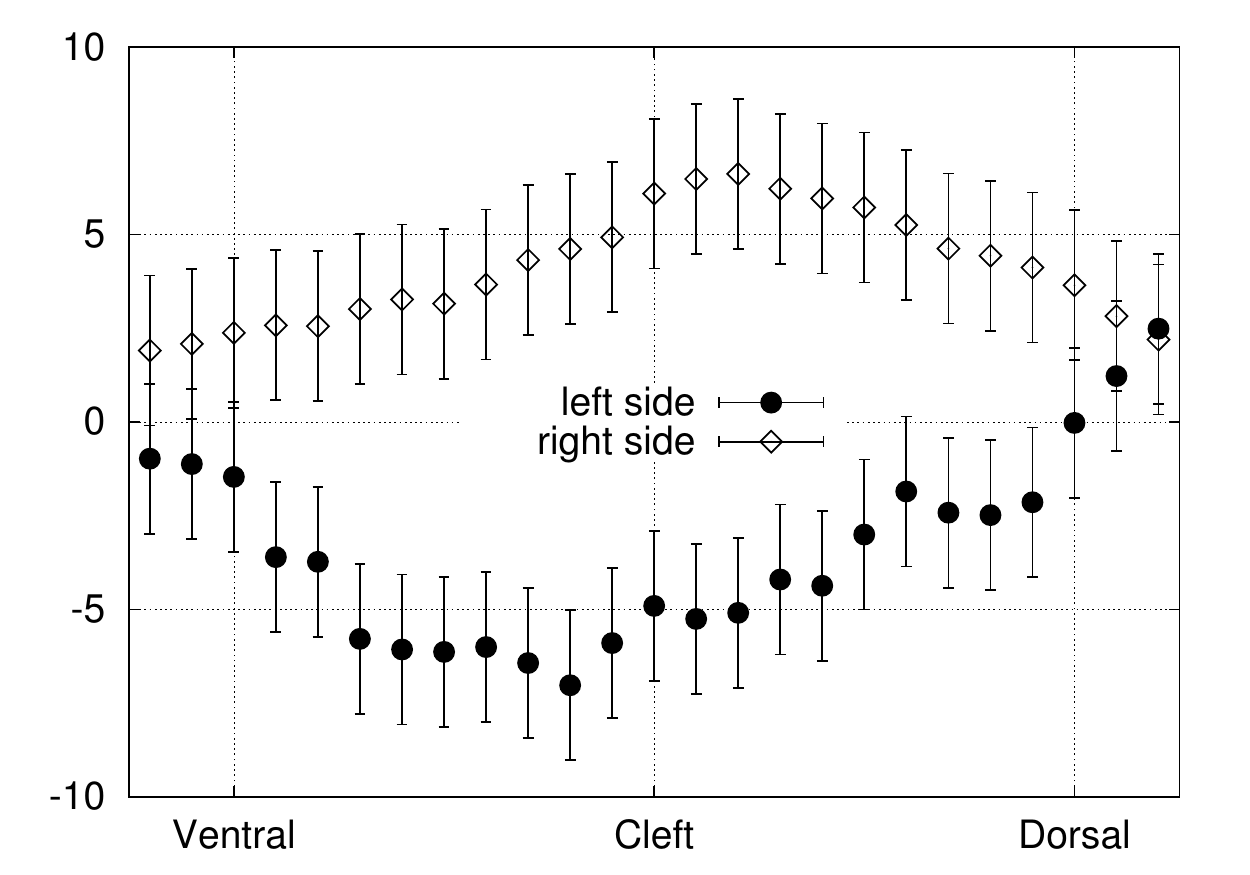}
\includegraphics[width=5.5cm]{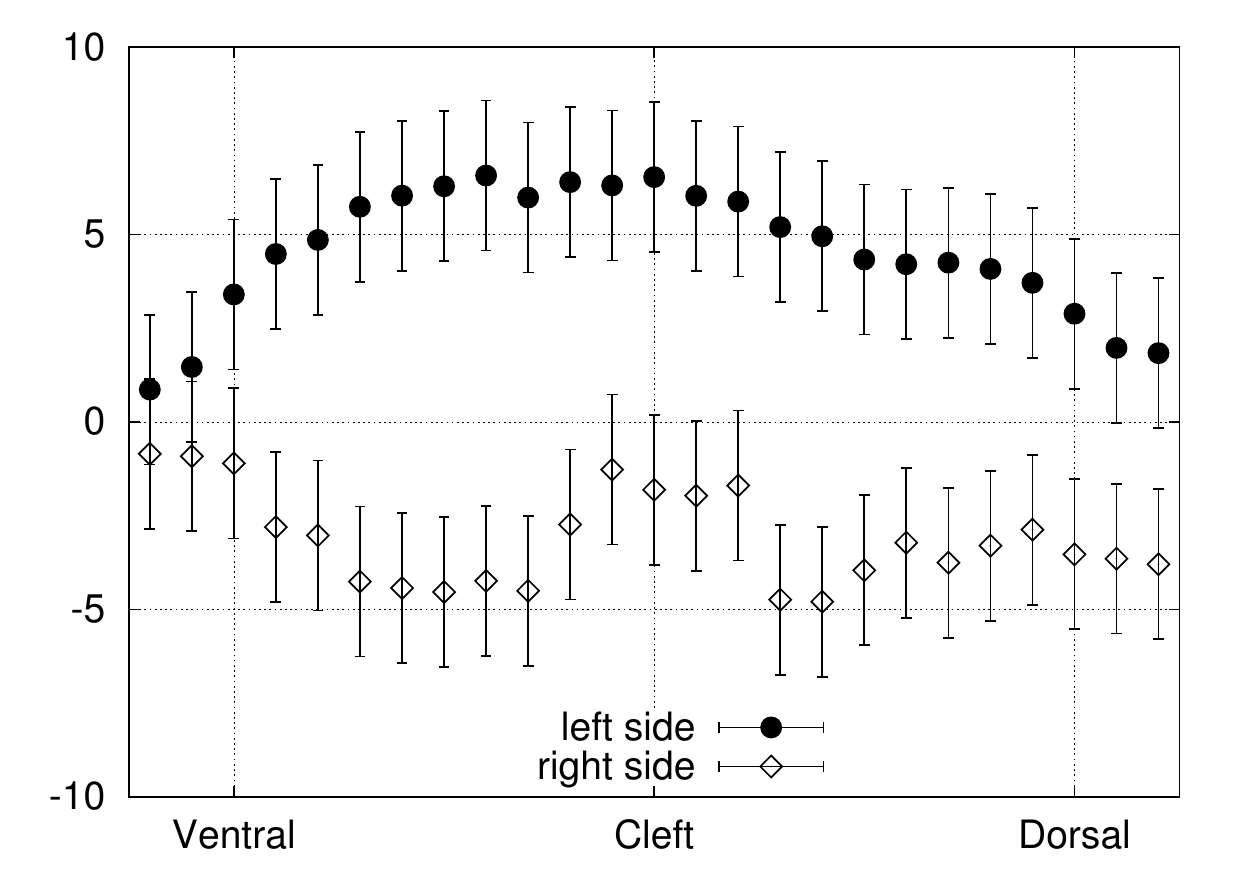}
\caption{Exp. 1 - Normal stress component (Pa), averaged over the most distal portion (15\%) of the left and right sides surfaces, plotted against the dorso-ventral axis. \textit{Top panel}: $t_{1/4}$. \textit{Bottom panel}: $t_{3/4}$.}
\label{f:t1_4_t3_4_Sn}
\end{figure}

The examination of instantaneous stress maps at mirroring time points (equivalent motion instants such as $t_{0}$ and $t_{1/2}$ or $t_{1/4}$ and $t_{3/4}$) is used to corroborate the consistency of the symmetrical distributions. Spatial symmetry between the left/right and ventral/dorsal sides of the hydrofoil is also verified to validate the stress maps. The stress patterns are drawn more explicitly by plotting the individual components along the relevant axes. The curves shown in figure \ref{f:t1_4_t3_4_Sn} represent the normal stress component (at $t_{1/4}$ and $t_{3/4}$) plotted against the dorso-ventral axis, averaged over the most distal portion (last 15\%) of the membrane. By comparing both panels of figure \ref{f:t1_4_t3_4_Sn}, we notice that as the fin is passing through mirroring sites, the normal stress on the left and right sides are mirrored. The absolute value of the pressure is maximal at the center of the foil and decreases towards the dorsal and ventral edges. The side which is leading the motion (the left side at $t_{1/4}$ and the right side at $t_{3/4}$) experiences a negative normal stress, which means that the stress is directed towards the surface. On the the trailing side, the positive normal stress points in the direction of the outward normal, as the fluid tends to hold the foil back.

\begin{figure}
\centering
\includegraphics[width=5.5cm]{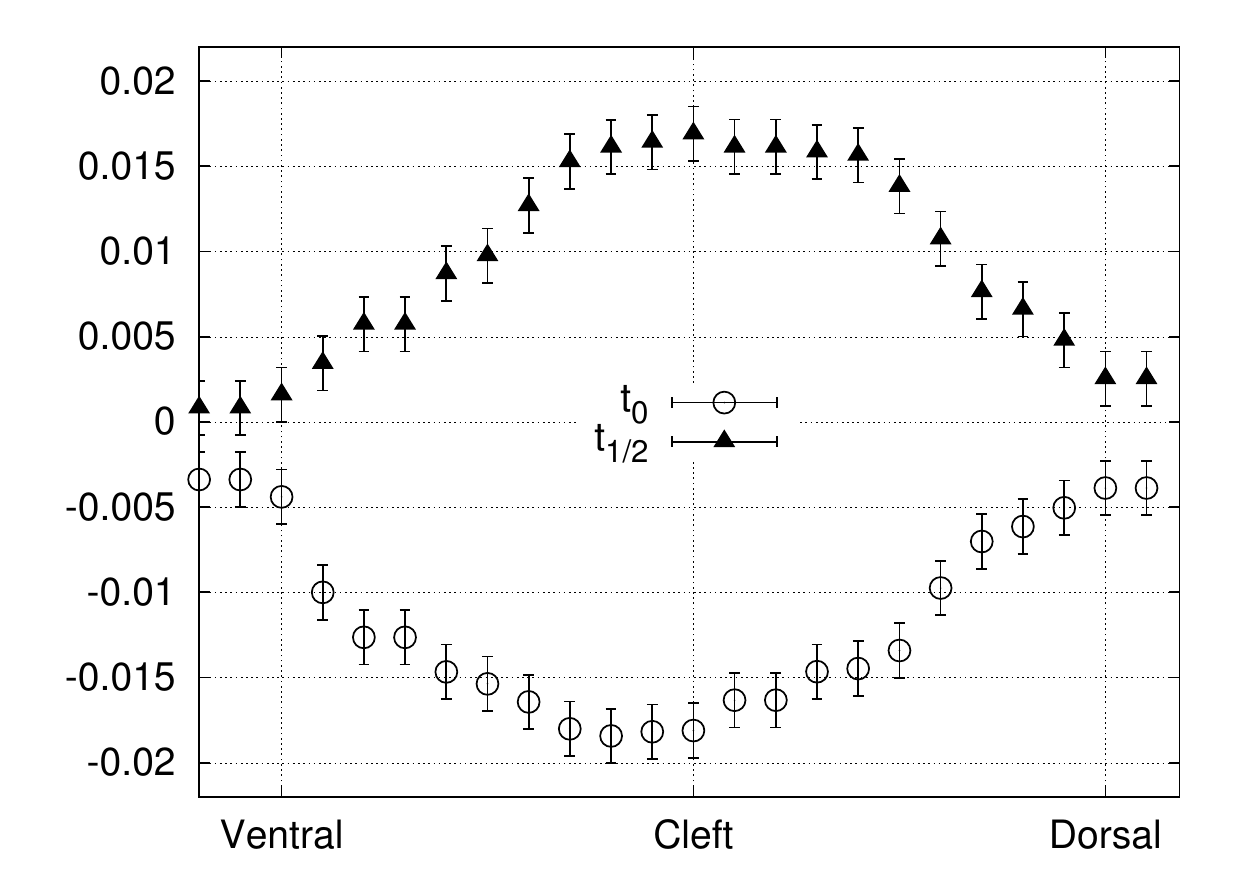}
\includegraphics[width=5.5cm]{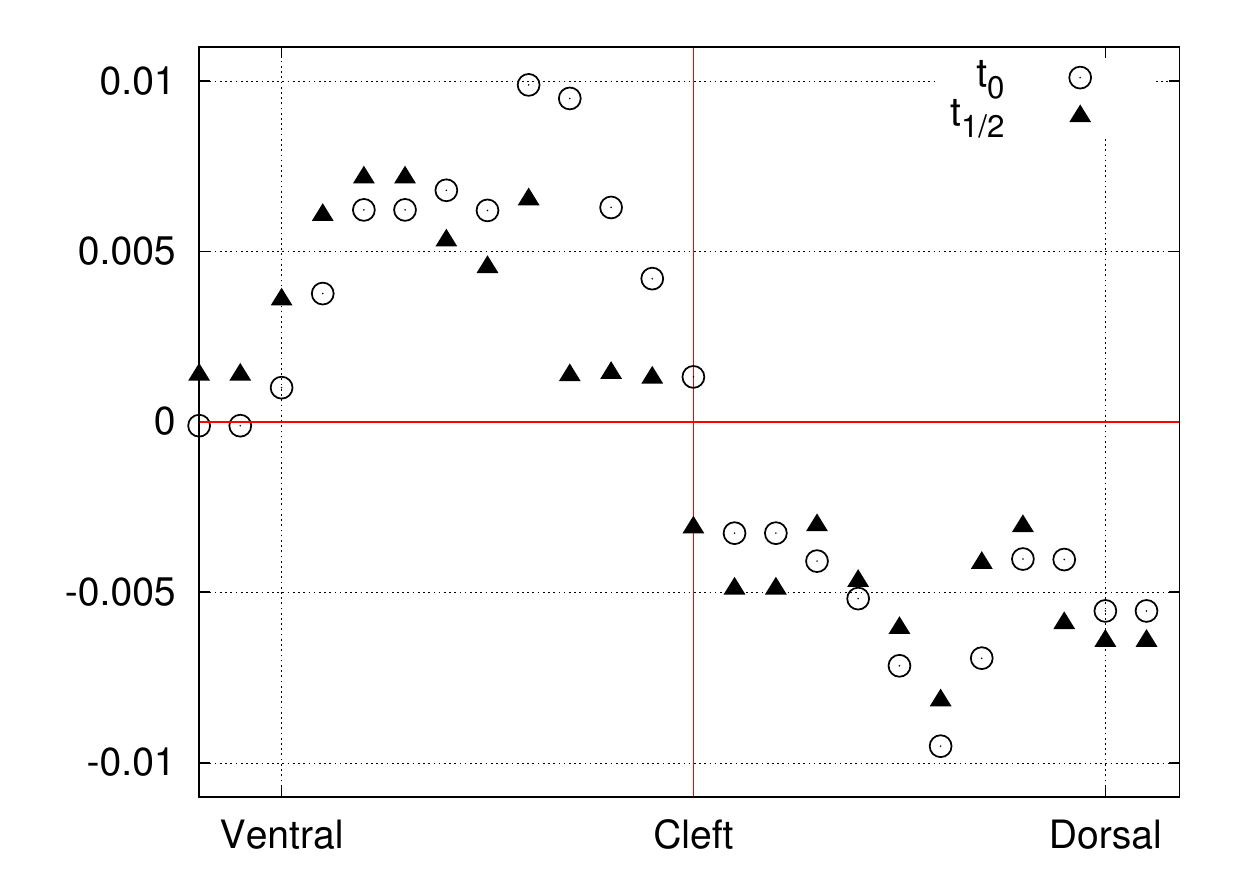}
\caption{Exp. 1 ($t_0$ and $t_{1/2}$) - Shear stresses (Pa) on the tip surface, plotted against the dorso-ventral axis. \textit{Top panel}: Left-right (transversal) stress component. \textit{Bottom panel}: Dorso-ventral stress component.}
\label{f:t0_t1_2_Slr_Sdv}
\end{figure}

In figure \ref{f:t0_t1_2_Slr_Sdv}, we consider the viscous shear stresses on the tip surface, at mirroring time points $t_{0}$ and $t_{1/2}$. At $t_{0}$, the fin travels towards its right side (positive $y$ values) and the tip surface experiences shear stress pointing in the left direction ($S_{lr}<0$), with a maximum absolute value in the center. The opposite trend is observed at $t_{1/2}$. As for the dorso-ventral shear stress, it is positive on the ventral portion of the tip, and negative on the dorsal part, implying that the stress is directed towards the center as the fluid tends to contract the fin tip surface. Naturally, this effect occurs for both $t_{0}$ and $t_{1/2}$ since the dorso-ventral axis is symmetric between the mirroring time frames.

Finally, the period-averaged normal stress map is illustrated in figure \ref{f:exp1_Sn_ave}. Three curves are shown, each one corresponding to a selected position along the proximo-distal axis of the fin: at 25\%, 50\% and 100\% of the total length (0.25$L$, 0.5$L$ and $L$).

\subsubsection{Experiment 2}

\begin{figure}
\centering
\includegraphics[width=5.5cm]{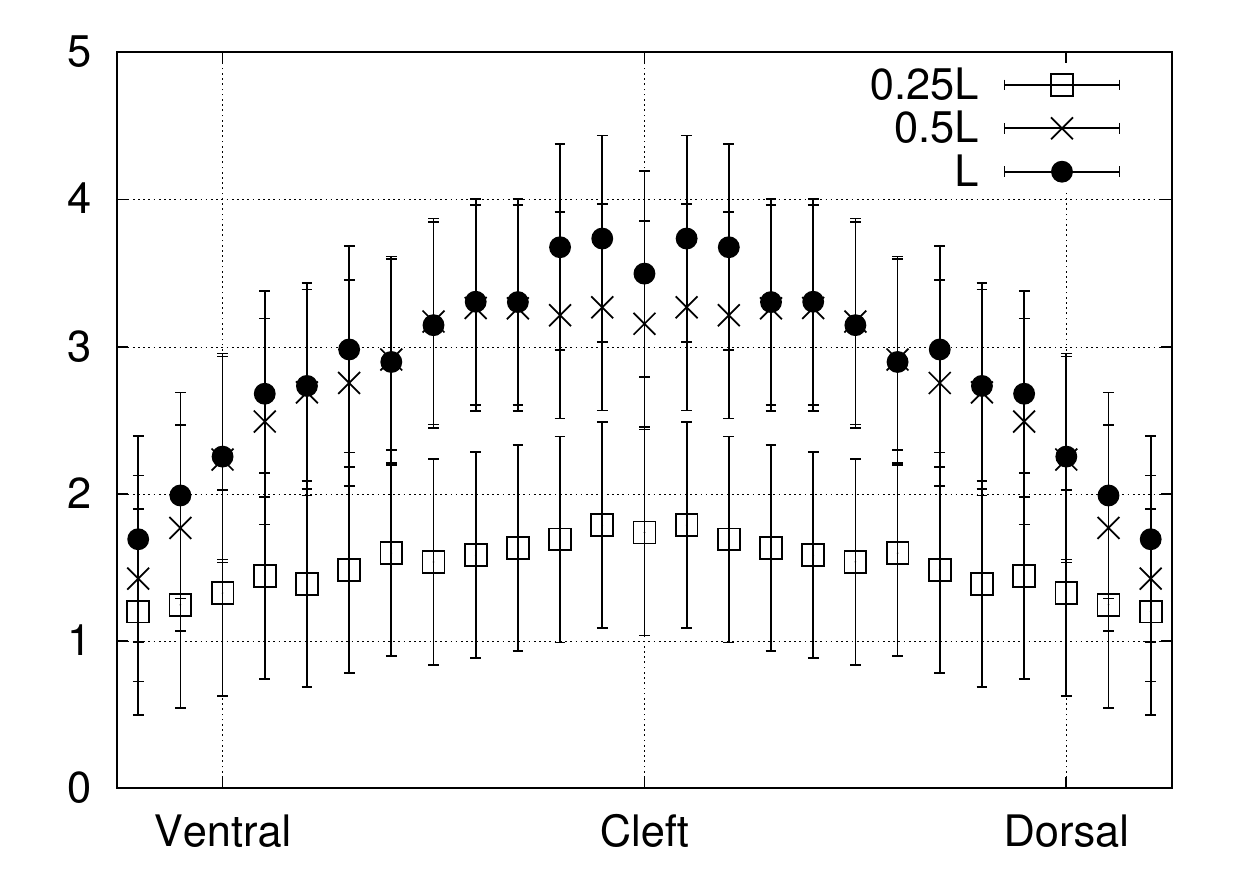}
\caption{Period-averaged absolute value of the normal stress component (Pa) on the sides surfaces (exp. 1), plotted against the dorso-ventral axis}
\label{f:exp1_Sn_ave}
\end{figure}

\begin{figure}
\centering
\includegraphics[width=5.5cm]{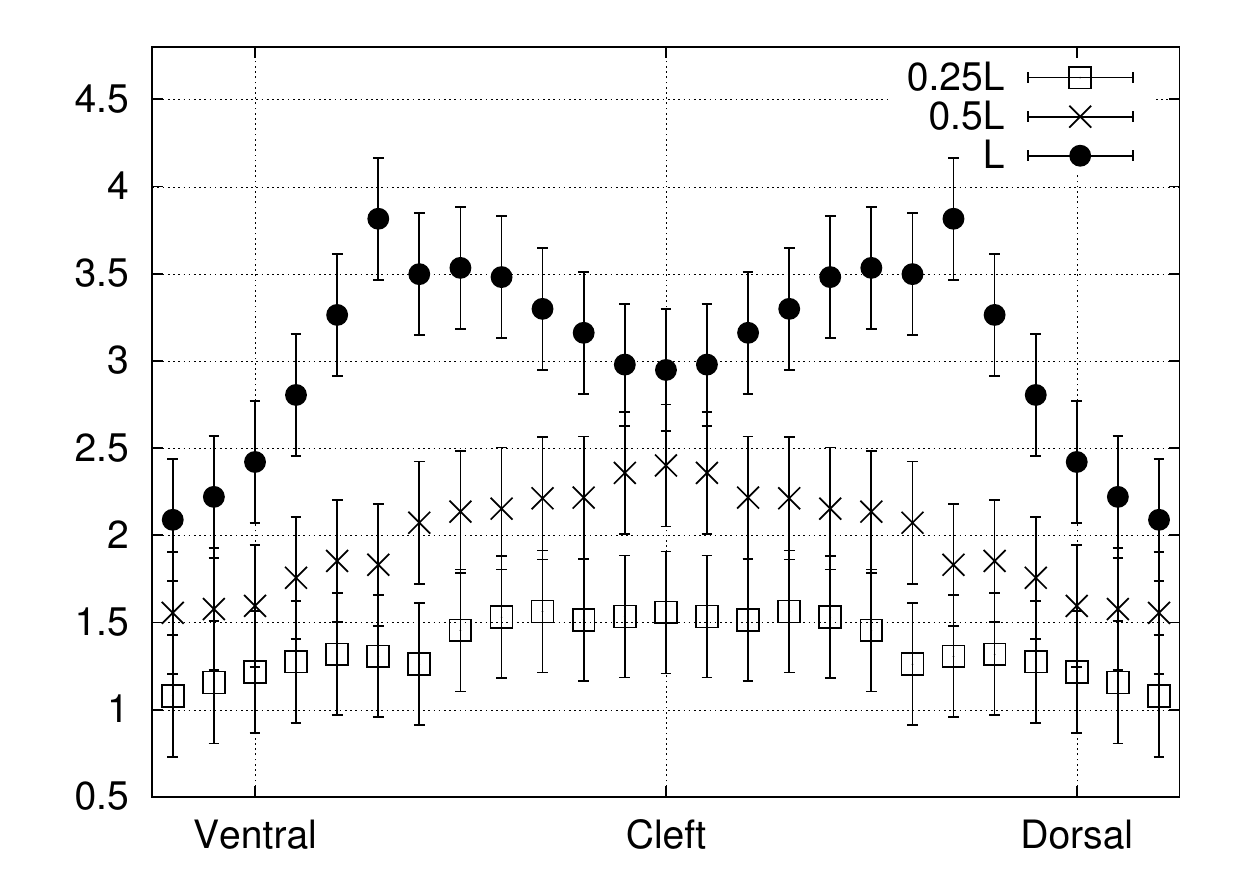}
\caption{Period-averaged absolute value of the normal stress component (Pa) on the sides surfaces (exp. 2), plotted against the dorso-ventral axis}
\label{f:exp2_Sn_ave}
\end{figure}

An interesting application of the hydrodynamic stress mapping method lies in the investigation of particular stress signatures or thresholds along the main axes of the hydrofoil, where the effects of varying specific flow parameters can be analyzed. The period-averaged normal stress distributions are compared between exp. 1 (figure \ref{f:exp1_Sn_ave}) and exp. 2 (figure \ref{f:exp2_Sn_ave}) to deduce how the hydrodynamic load is changing over the fin surface depending on the flow conditions. The magnitude of the normal stress at 50\% of the fin length (0.5$L$) is reduced as a consequence of increasing the Reynolds number and reducing the Strouhal number by applying an external flow (exp. 2). Besides, at the distal location on the foil ($L$), the curve goes from a single peak for exp. 1 to a curve with two maxima for exp. 2, with similar maximum magnitudes. Therefore, the presence of an external advecting flow, combined with slightly larger flapping amplitude, shifts the stress curves to lower values at specific positions along the fin (about half-way through the proximo-distal axis) and modifies the peaks profile on the most distal surface. These results support the practicality of the stress mapping method as a powerful tool for analyzing the distribution of fluid forces on submerged and deforming objects.

\subsection{Validations}
\label{sst:results_validation}

\begin{figure}
\centering
\includegraphics[width=8cm, trim={0.5cm 0.5cm 0cm 0.5cm}]{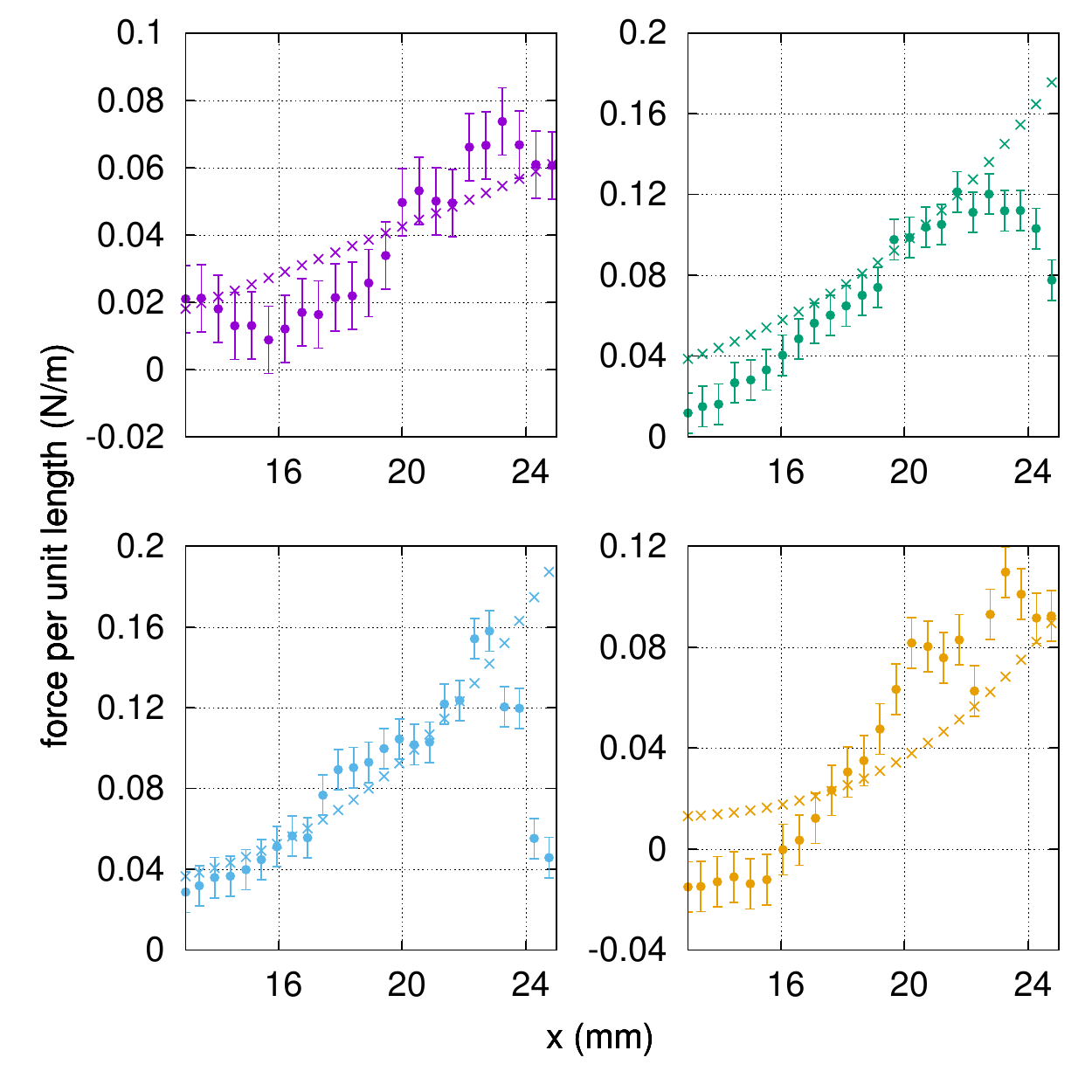}
\caption{Force per unit length along the proximo-distal axis of the fin (exp. 1), from the PTV-based stress distributions (full circles) and the Euler-Bernoulli load calculation (crosses), at $t_{0}$, $t_{1/8}$, $t_{1/4}$ and $t_{3/8}$ from top left to bottom right, respectively. Color code follows that of figure \ref{f:foil_kinematics}.}
\label{f:validation}
\end{figure}

The motion instants $t_{0}$, $t_{1/8}$, $t_{1/4}$ and $t_{3/8}$ from exp. 1 (covering half a period) were selected for the force results validation. The parametrized deflection of the foil (equation \ref{eq:deflection_fit}) was used to compute the appropriate derivatives in equation \ref{eq:euler_bernoulli}, and to provide an estimate of the transverse force per unit length (pointing upwards) along the proximo-distal axis of the fin. The results are presented in figure \ref{f:validation}, and compared with the force distributions obtained by integrating the PTV-based normal stress distributions over both sides of the hydrofoil. The results are in good agreement, which validates the accuracy of the hydrodynamic stress maps obtained from the velocity fields. The force drop at the tip of the fin for $t_{1/8}$ and $t_{1/4}$, captured by the PTV-based stresses, is not well reproduced by the Euler-Bernoulli model. We attribute this small discrepancy to the larger tip deflections at these specific time points, where the Euler-Bernoulli formalism is less applicable by definition. Moreover, the quality of the fit (equation \ref{eq:deflection_fit}) was slightly decreased at the tip of the fin for these time points, where the deflection was overestimated by the fitted function. The main contribution to the total transverse load is the elastic restoring force, which is of the order of 0.1 N/m, followed by the inertial force, of the order of 0.01 N/m. The linear fluid damping and the structural damping contributions are negligible in comparison, of the order of 0.001 N/m. In brief, the force distributions obtained from the Euler-Bernoulli deflection analysis offer a validation of the PTV-based normal stresses.

Additionally, a control volume analysis was used to validate the consistency of the instantaneous force data. The total forces acting on the fin in the $x$ and $y$ directions ($F_x$ and $F_y$) were calculated from equation \ref{eq:control_volume}, for the eight selected time points of exp. 1. Good agreement was found between the control volume forces and the values obtained by integrating the stress distributions on the fin surfaces within the uncertainty of the pressure determination corresponding to an uncertainty in the force of $\sigma_F = 0.4$mN. Both approaches describe a net force in the negative $x$ direction with a maximal magnitude of $\sim$0.75 mN at mirroring instants $t_{1/4}$ and $t_{3/4}$ (where the hydrofoil tip excursion is maximal), implying that the foil experiences a net thrust as it is propelled upstream. Both methods also capture a net force in the positive $y$ direction in the first half of the period (instants $t_{1/8}$ to $t_{3/8}$) and in the negative $y$ direction in the second half of the period (instants $t_{5/8}$ to $t_{7/8}$), with a maximal magnitude of $\sim$1.5 mN. The estimation of the streamwise force $F_x$ from the control volume method offers poorer results that the vertical force profile $F_y$.  This is not surprising, as the control volume approach is typically employed on a larger domain containing all the fluid affected by the motion of the solid object, which is restricted in our case by the measurement domain (50$\times$50$\times$20 mm\textsuperscript{3}). Consequently, the momentum balance inside the control volume predicts a fallacious positive peak of $F_x$ in the middle of the period, whereas the stress distributions integrated on the hydrofoil indicate a zero net force in the $x$ direction at that instant ($t_{1/2}$). The vertical force is better captured, due to the fact that the fluid motion in the $y$ direction is more thoroughly included inside the control volume. We conclude that the time-evolution of the PTV-based stress distributions are consistent with the fluid forces obtained from a control volume analysis, except for a certain offset of the magnitude in the $x$ direction due to the limited field of view in this case.

\subsection{Uncertainties}
\label{sst:results_uncertainty}

From equation \ref{eq:error_pressure_statistical}, it is possible to provide a raw estimate of the pressure uncertainty, using the parameters specific to our experiments. With an approximate integration path length (covering half the fluid domain) of 30 nodes in the $x$ and $y$ directions and 10 nodes in the $z$ direction, we get an approximate pressure error of 1 Pa (after averaging over the $x$, $y$, $z$ directions).

A more detailed analysis is rendered possible by the use of equation \ref{eq:error_pressure}. In the case of exp. 1, the error on the material acceleration arising from the Lagrangian path reconstruction inaccuracy (the second term in equation \ref{eq:error_material_acceleration} which carries the spatial dependence) was calculated everywhere in the 3D domain. Spatial average over the whole volume and temporal average over a complete period (using the 8 time frames $t_{0,\ldots,7/8}$) yielded the following values:

$$ \langle (\vec{\sigma}_u \cdot \vec{\nabla})u_x \rangle=0.029 m/s^2$$
$$ \langle (\vec{\sigma}_u \cdot \vec{\nabla})u_y \rangle=0.023 m/s^2$$
$$ \langle (\vec{\sigma}_u \cdot \vec{\nabla})u_z \rangle=0.040 m/s^2$$

The larger value obtained for the $z$ component contributes to larger uncertainties accumulating in that direction. In comparison, the first term of equation \ref{eq:error_material_acceleration}, which contains the material acceleration uncertainty propagated from the PTV errors, produces values at least one order of magnitude larger:

$$\langle \frac{\sqrt{2}\sigma_{u_x}}{\Delta t} \rangle=0.23 m/s^2$$
$$\langle \frac{\sqrt{2}\sigma_{u_y}}{\Delta t} \rangle=0.23 m/s^2$$
$$\langle \frac{\sqrt{2}\sigma_{u_z}}{\Delta t} \rangle=2.04  m/s^2$$

\begin{figure}
\centering
\includegraphics[width=5.5cm]{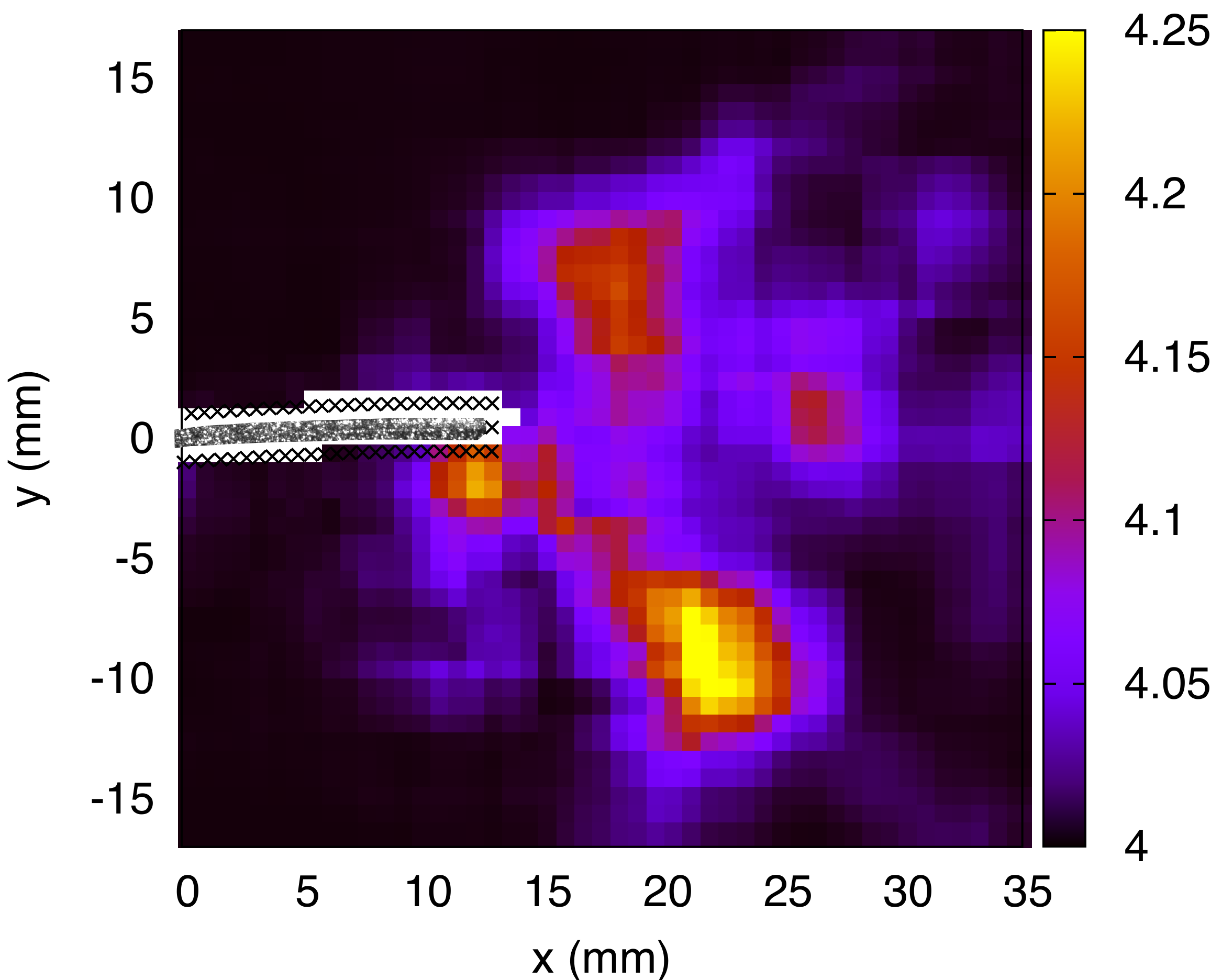}
\includegraphics[width=5.5cm, trim={0cm 1.5cm 1cm 1.5cm}]{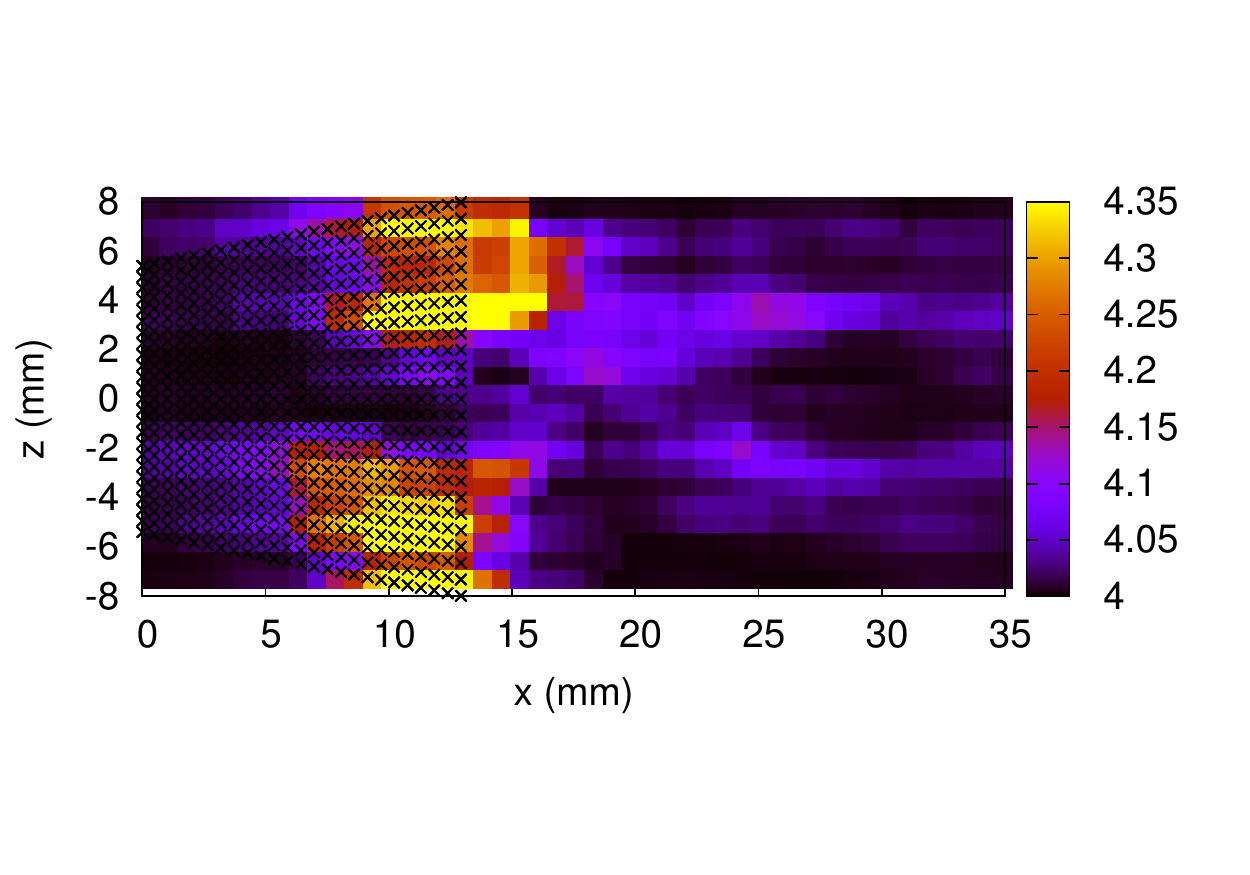}
\caption{Pressure uncertainty in Pa (exp. 1, $t_0$). \textit{Top panel}: vertical midplane. \textit{Bottom panel}: horizontal plane at $y$=-5.275 mm. }
\label{f:exp1_t0_Perror}
\end{figure}

\begin{figure}
\centering
\includegraphics[width=5.5cm]{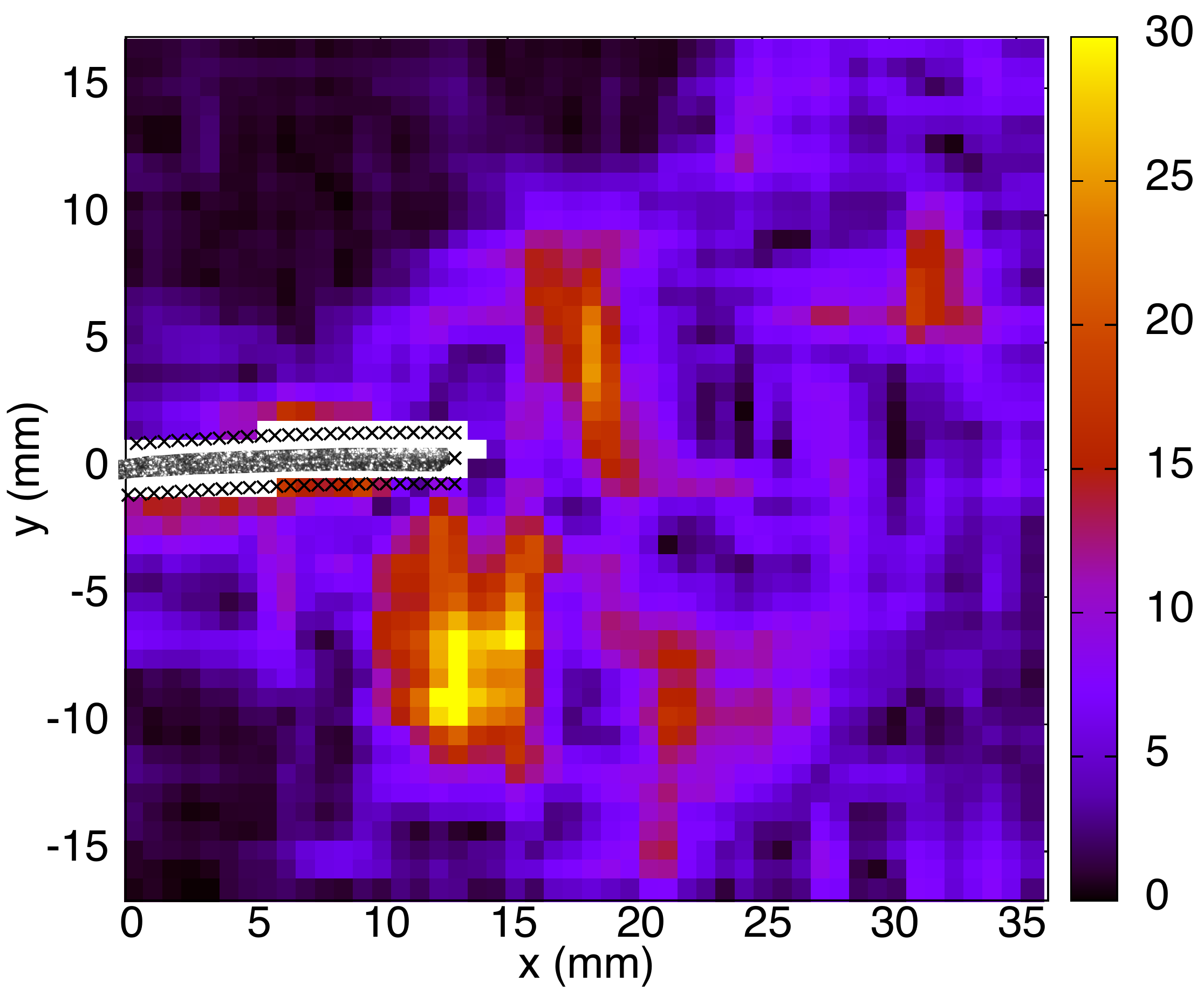}
\includegraphics[width=5.5cm, trim={0cm 1.5cm 1cm 1.5cm}]{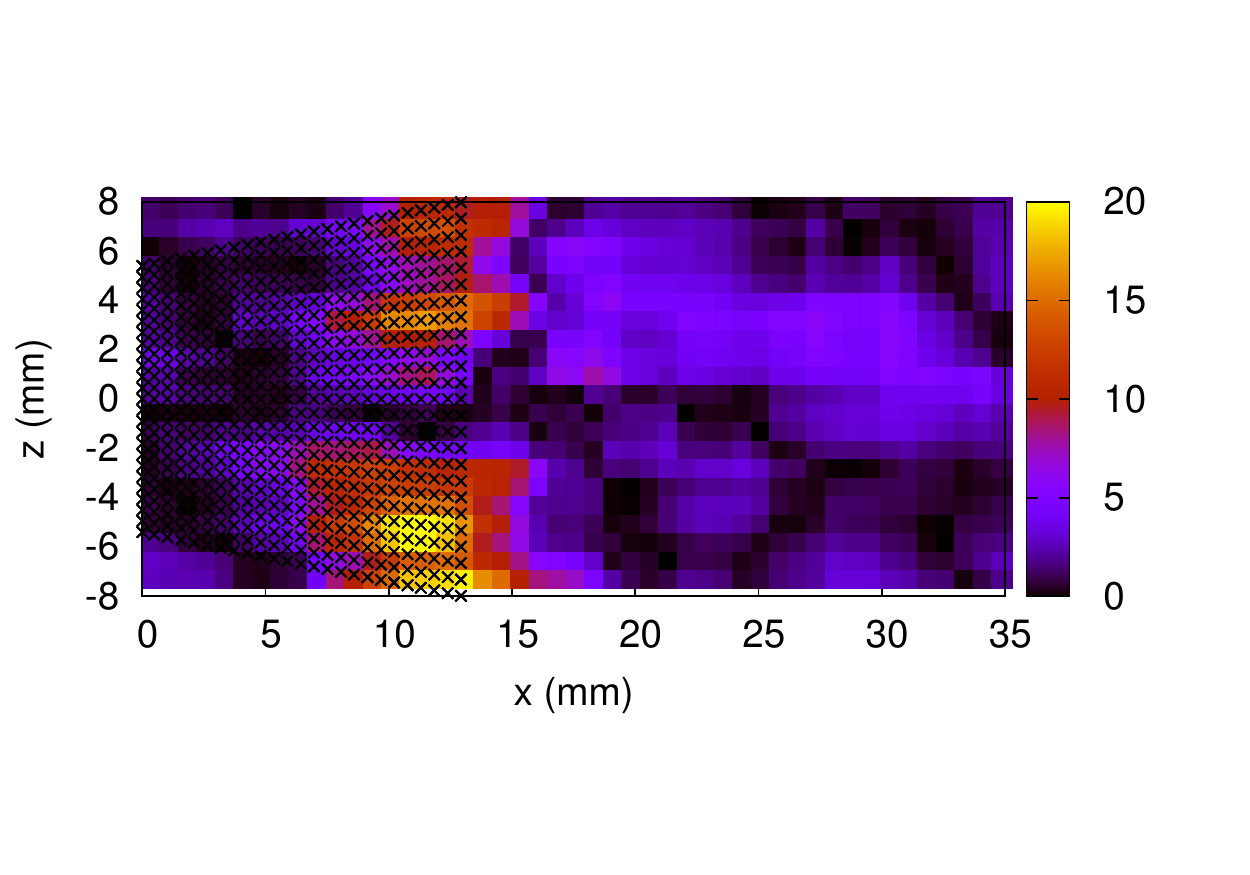}
\caption{Fluid vorticity in s\textsuperscript{-1} (exp. 1, $t_0$). \textit{Top panel}: Vorticity magnitude in the vertical midplane. \textit{Bottom panel}: Vorticity $x$-component (in absolute value, $\abs{\omega_x }$) in a horizontal plane ($y$=-5.275 mm). }
\label{f:exp1_t0_rotational}
\end{figure}

We thus conclude that the error propagated from the velocity field constitutes the main source of error in the pressure calculation, whereas the contribution from the inaccuracy of the Lagrangian path reconstruction is almost negligible. The resulting pressure uncertainty ($\sigma_p$), averaged over the 3D domain and a complete motion cycle, is equal to 4.03 Pa. The spatial dependence of $\sigma_p$ is shown in figure \ref{f:exp1_t0_Perror} (for $t_0$, exp. 1). The fluid vorticity ($\abs{\vec{\omega}}$) and its streamwise component (in absolute value, $\abs{\omega_x}$) are illustrated in figure \ref{f:exp1_t0_rotational}. The pressure uncertainty is found to be spatially correlated to the vorticity, in particular, the error becomes larger vis-a-vis the tip corners, where high vorticity around the $\hat{x}$ axis is found. This correlation is expected since the spatial dependence of $\sigma_p$ is carried by the term associated to errors in the pseudo-tracking method. Thus, it is no surprise that the regions where the particles follow paths with strong curvature (high vorticity) are the regions where the Lagrangian path reconstruction, which assumes a linear displacement between subsequent velocity fields, performs more poorly.

Including the reduction factor due to spatial and temporal averaging, the final uncertainties on the normal stress in exp. 1 are $\sigma_p \simeq$ 2 Pa for the instantaneous distributions (figure \ref{f:t1_4_t3_4_Sn}) and 0.7 Pa for the period-average (figure \ref{f:exp1_Sn_ave}). The uncertainties on the viscous shear stresses (acting on the tip, figure \ref{f:t0_t1_2_Slr_Sdv}) are $\sigma_{\tau_{yx}} \simeq$ 1.6 mPa for the left-right component and $\sigma_{\tau_{zx}} \simeq$ 10 mPa for the dorso-ventral component. However, the dorso-ventral symmetry and the temporal symmetry in the shear stress curves suggest on the contrary that these stress distributions are well resolved within an uncertainty range that does not exceed 3 mPa. We interpret this as an indication that the velocity errors stated in section \ref{sst:uncertainty_calculation} are slightly overestimated, as they do not include the appropriate error reduction associated to the filtering and smoothing steps. For that reason, the error bars were exceptionally omitted on that specific graph (figure \ref{f:t0_t1_2_Slr_Sdv}, lower panel). For exp. 2, the error bars in figure \ref{f:exp2_Sn_ave} correspond to $\sigma_p \simeq$ 0.35 Pa.

Finally, the error bars of figure \ref{f:validation} were calculated based on $\sigma_p$, the integration path length (the average fin width) and the number of grid points along that path (25 on each side of the fin).

\section{Discussion and outlook}
\label{st:discussion}

In the present study of a prototypical fin-like foil, we tested the limits of hydrodynamic stress calculations based on a pressure gradient multi-path integration scheme, which we applied to a time series of volumetric PTV measurements of an unsteady vortical flow. This method was shown to produce well resolved stress curves along the biologically relevant axes of the synthetic fin, allowing to gain precise information about the interaction between the deforming surface and its direct fluid environment. The non-intrusive approach of hydrodynamic stress mapping constitutes a promising path to deepen our understanding of the relation between the morphometric and elastic properties of hydrofoils and the fluid forces which they experience. We foresee numerous applications to this experimental methodology, in domains such as the engineering of underwater propulsive foils or the investigation of hydrodynamic forces in aquatic animal locomotion. Propulsive foils involved in marine robotic systems would greatly benefit from the employment of small scaled models for which the distribution of hydrodynamic stresses and their time evolution could be described everywhere on the surface, without the need for intrusive mechanical transducers, thus guiding the structural design with precise knowledge of the local mechanical loads. In a parallel train of thought, mapping the stress patterns on the surface of biomimetic synthetic fins can provide new insights about the mechanical constraints under which real fins have evolved to attain their exact morphology.

\begin{acknowledgements}
This work was funded by the Swiss National Science Foundation (SNF) via a Sinergia research grant as well as a UZH Forschungskredit Candoc grant. We are grateful for interdisciplinary discussions with Tinri Aegerter, Anna Ja\'zwi\'nska, Ivica Kicic, Sahil Puri and Siddhartha Verma. We are very thankful to J.O. Dabiri \textit{et al.} for making the \textit{queen2} pressure algorithm available \cite{Dabiri2014}.
\end{acknowledgements}

 \section*{Conflict of interest}
The authors declare that they have no conflict of interest.

\bibliographystyle{spphys}
\bibliography{stress_fin_6_bib}

\end{document}